# Unconventional Cognitive Intelligent Robotic Control: Quantum Soft Computing Approach in Human Being Emotion Estimation - QCOptKBTM Toolkit Application

Ulyanov Sergey V.*, Kurawaki I. **, Ulyanov Viktor S.† and Hagiwara T. ***

*Institute of System Analysis and Management, Dubna State University
* Meshcheryakov Laboratory of Information Technologies, Joint Institute for Nuclear Research (JINR)
†Department of Information Technologies, Moscow State University of Geodesy and Cartography (MIIGAiK)
**, *** Yamaha Motor Co. Ltd., Automotive operations Dpt.
*Email: srg.v.ulyanov@gmail.com
†Email: ulyanovik@gmail.com
** Email: kurawakii@yamaha-motor.co.jp
*** Email: hagiwarat@yamaha-motor.co.jp

**Abstract**

Strategy of intelligent cognitive control systems based on quantum and soft computing presented. Quantum self-organization knowledge base synergetic effect extracted from intelligent fuzzy controller's imperfect knowledge bases described. That technology improved of robustness of intelligent cognitive control systems in hazard control situations described with the cognitive neuro-interface and different types of robot cooperation. Examples demonstrated the introduction of quantum fuzzy inference gate design as prepared programmable algorithmic solution for board embedded control systems. The possibility of neuro-interface application based on cognitive helmet with quantum fuzzy controller for driving of the vehicle is shown.

## Introduction

The article shows the possibility of quantum / soft computing optimizers of knowledge bases (QSCOptKB™) as the toolkit of quantum deep machine learning technology implementation in the solution's search of intelligent cognitive control tasks applied the cognitive helmet as neuro-interface. In particular case, the aim of this article is to demonstrate the possibility of classifying the mental states of a human being operator in on line with knowledge extraction from electroencephalograms based on SCOptKB™ and QCOptKB™ sophisticated toolkit. Application of soft computing technologies to identify objective indicators of the psychophysiological state of an examined person described. The role and necessity of applying intelligent information technologies development based on computational intelligence toolkits in the task of objective estimation of a general psychophysical state of a human being operator shown. Developed information technology examined with examples emotion state estimation of human being operator in "Human – Robot Interaction" interface design and background of the knowledge bases design for intelligent robot of service use is it. Application of cognitive intelligent control in navigation of autonomous robot for avoidance of obstacles based on QFI demonstrated.

## 1. Tasks of hybrid cognitive and intelligent quantum control

A number of studies [1-7] showed the possibility of development a simplified mathematical model of emotions. But due to physical limitations, the trade-off of informational boundaries on the applicability of the developed model also have a significant influence on the correctness of description and reliability of the extracted knowledge from the imperfect mathematical model. In intelligent control system (ICS) theory, one effective approaches to the risk decreasing of decision-making is the development of robust ICS structures with corresponding knowledge bases (KBs).

The problems of physical limitations and information boundaries solved by the possibility of forming KB with the required level of robustness in the design process of ICS by extracting knowledge and valuable information from the dynamic behavior of the model of the physical control object [8].

*Figure 1* demonstrates general structure of hybrid cognitive intelligent control system.

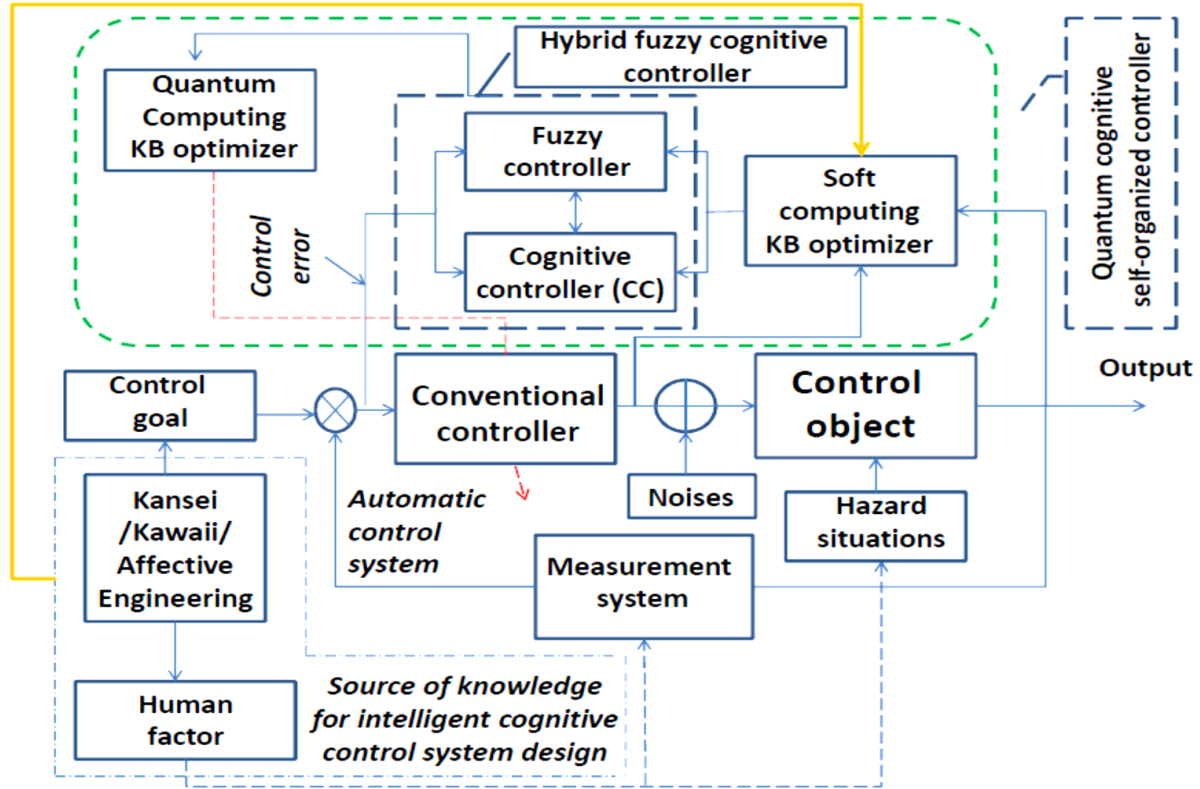

*Figure 1: Structure of hybrid intelligent cognitive control system based on quantum soft computing.*

The structure based on fuzzy and cognitive controllers, includes quantum fuzzy inference (QFI) with quantum genetic algorithm in Box "Quantum computing KB optimizer" and are the background of quantum cognitive self-organized controller (see, in details [8-10]).

The main problem of cognitive intelligent control system (presented in *Fig. 1*) is to design optimal robust control with minimal loss of value work and minimum of required initial information on external environments.

Let us consider briefly (see in details [8-11]) the solution of this problem using information-thermodynamic approach.

Let us consider a network of loosely coupled groups of robots working together to solve tasks that go beyond individual capabilities. Different nodes of such a system have a different intelligent level (knowledge, algorithms, and computational bases) and various information resources in designing. Each node should be able to modify its behavior depending on the circumstances, as well as to plan its communication and cooperation strategies with other nodes. Here the indicators of the level of cooperation are the nature of the distribution of tasks, the unification of various information resources and, of course, the possibility of solving a common problem in a given time.

## 2. Quantum algorithm of knowledge self-organization

A quantum algorithm (QA) model of ICS self-organization, proposed in [11], is based on the principles of minimum information entropy (in the "intelligent" state of control signals) and a generalized thermodynamic measure of entropy production (in the system "control object + controller"). The main result of the application of the self-organization process is the acquisition of the necessary level of robustness and the flexibility (adaptability) of the reproducible structure. It is noted that the property of robustness (by its physical nature) acts as an integral part of self-organization, and the required level of robustness of ICS is achieved by fulfilling the principle of minimum production of generalized entropy, which was noted above.

The principle of minimum entropy production [12-15] in control object and control system [16] serves as the physical principle of optimal functioning with a minimum consumption of useful work and underlies the development of robust ICS. This statement based on the fact that for the general case of controlling dynamic objects, the optimal solution is to the finite variation problem of determining the maximum of the useful work *W* is equivalent, according to [17], to the solution of the finite variation problem of finding the minimum of the entropy production *S*.

Therefore, the developed QA model used principle of minimum informational entropy guarantees the *necessary* condition for self-organization — the minimum of the required initial information in the teaching signals; the thermodynamic criterion of the minimum of a new measure of generalized entropy production provides a *sufficient* condition for self-organization - the robustness of control processes with a minimum consumption of useful resource.

More significant is the fact that the average amount of work done by dissipation force

$$\frac{\langle W_{diss} \rangle}{kT} = S_{KL}(P_F||P_B),$$

i.e., the work of dissipation forces is determined by the Kullback-Leibler divergence for probability distributions $P_F, P_B$. Note that the left side of this relation represents physically thermal energy, and the right side defines purely information about the system. A similar relationship exists between the work produced by the forces of dissipation and the difference between generalized Renyi divergences [18].

*Figure 2* illustrates the QA structure of self-organization (QASO) in design process of robust intelligent PID-controller with application of quantum fuzzy inference.

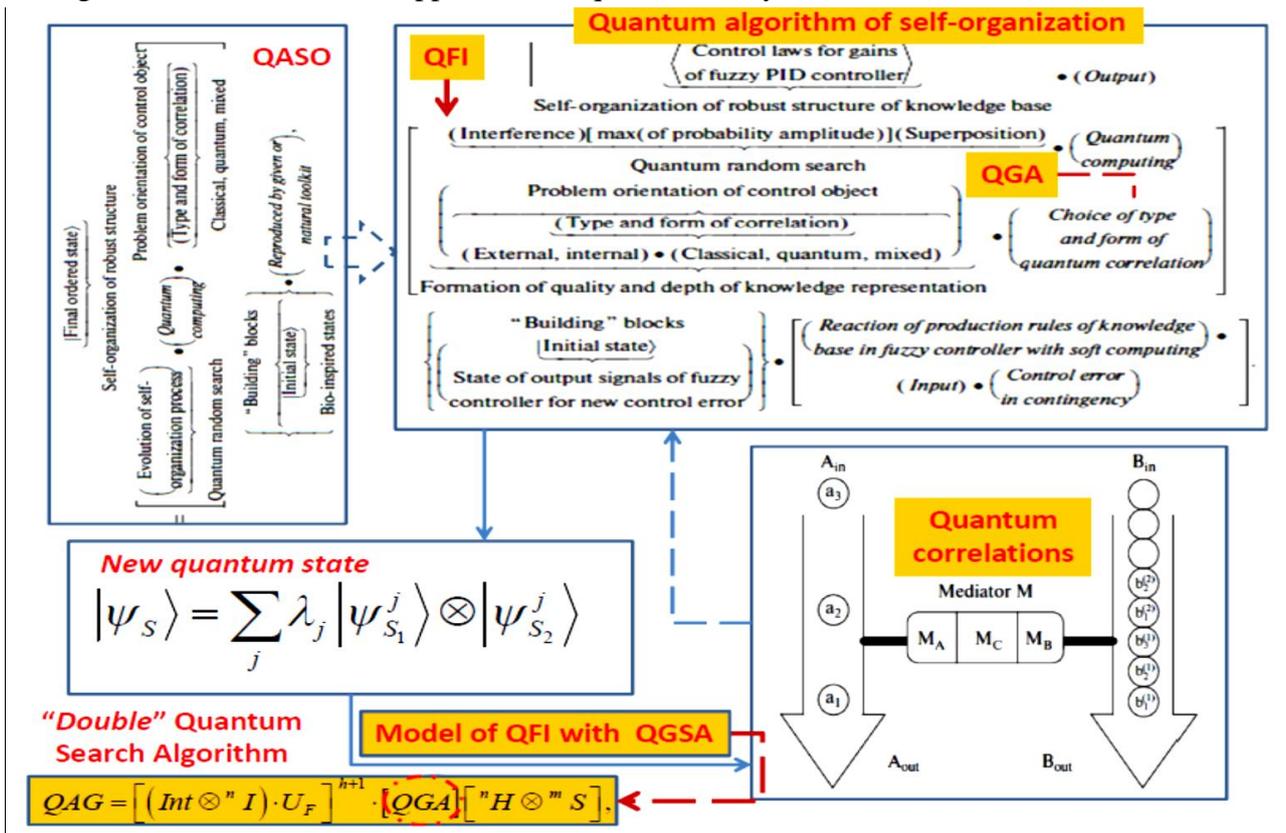

*Figure 2: Quantum algorithm of self-organization based on quantum fuzzy inference and quantum genetic algorithm.*

Structure of QFI include the quantum genetic algorithm for choice the optimal quantum correlation type between PID-controller coefficient gains in temporal schedule.

Thus, substituting the relations between the information and the extracted free energy and work in [10], we obtain the conclusion that the robustness of the intelligent control system can increased by producing the entropy of the cognitive controller. The optimal cognitive controller reduces the loss of useful resource of the control object, and negentropy of the cognitive regulator reduces the requirements for minimum initial information to achieve robustness.

Therefore, the extracted information, based on knowledge (in the knowledge base of the cognitive controller), allows to get an additional resource for useful work, which is equivalent to the appearance of a targeted action on the control object to guarantee the achievement of the control goal in unpredicted situations, for example, with BELBIC controller [19,20].

### 2.1. Problems in quantum intelligent control systems design

Modern control objects are complex dynamic systems that characterized by information uncertainty of model structures and control goals, a high degree of freedom and essential nonlinearities, instability, distributed sensors and actuators, high level of noise, abrupt jump changes in structure and dynamics, and so on. It is the typical information resources of unpredicted control situations. The structure design of robust advanced control systems for unpredicted control situations is the corner stone of modern control theory and systems. The degree to which a control system deals successfully with above difficulties depends on the *intelligent* level of advanced control system.

In Step I of developed design technology, we focus the main attention on the description of particular results of KB design and simulating intelligent control systems with essentially nonlinear CO with a randomly time-dependent structure and control goals. In this case, the aim of this step is to determine the robustness levels of control processes that ensure the required reliability and accuracy indices under the conditions of uncertainty of the information employed in decision-making (learning situations).

For Step II, the description of the strategy of robust structure's design of an intelligent control system based on the technologies of quantum and soft computing given. The developed strategy allows one to improve the robustness level of fuzzy controllers under the specified unpredicted or weakly formalized factors for the sake of forming and using new types of self-organization processes in the robust KB with the help of the quantum computing methodology. A particular solution of a given problem obtained by introducing a generalization of decision-making strategies in models of fuzzy inference in the form of a new quantum fuzzy inference (QFI) on a finite set of fuzzy controllers designed in advance [21].

The basis for the development of control systems is the proportional-integral-differentiating (PID) controller, which used in 70% of industrial automation, but often does not cope with the control task and works very poorly in unforeseen situations. Fuzzy controllers allow to partially expand the scope of PID controllers by adding production logic rules and partially adapt the system. The combined use of genetic algorithms (GA) and a fuzzy neural network made it possible fully adapt the system, but it takes time to train such a system, which is critical in emergency and unforeseen situations. Modeling the optimal training signal makes it possible to create partial self-organization in the system due to the formation of optimal trajectories of the gain of the PID controller. The application of quantum computing and, as a particular example, QFI allows increasing robustness without spending a temporary resource in online [21,22].

*Figure 3* shows the ICS structure with the combination of several fuzzy regulators and the quantum fuzzy controller.

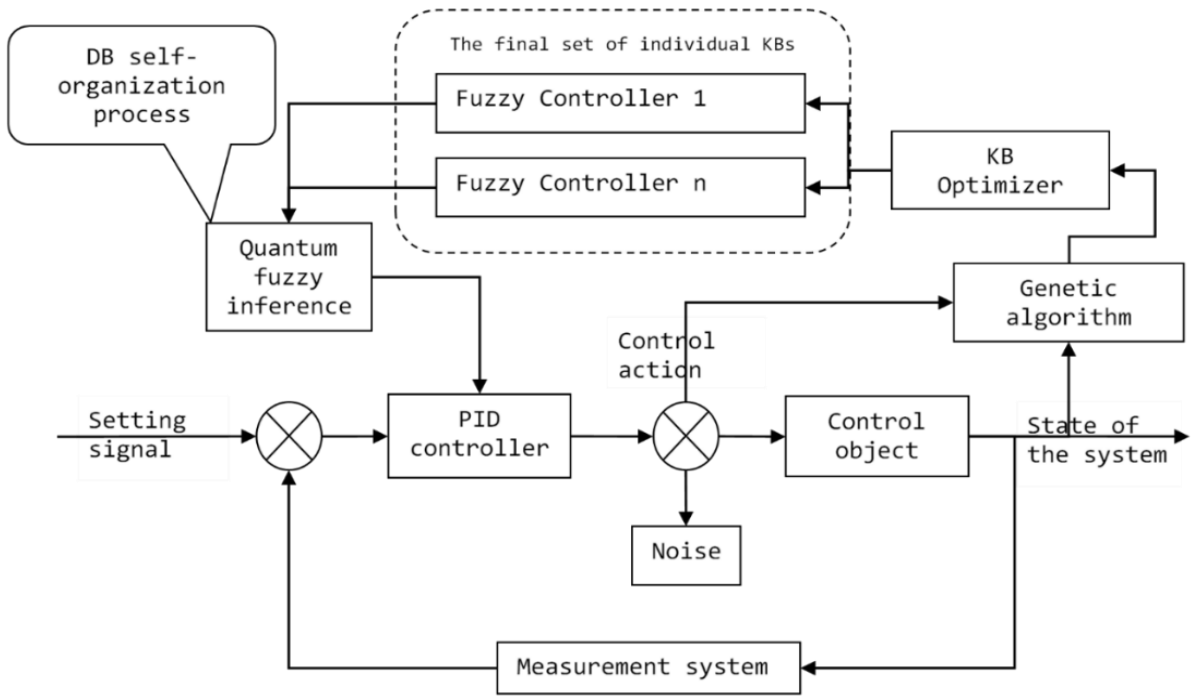

*Figure 3: Intelligent control system including quantum fuzzy inference.*

The main problem in the development and design of this structure that it is very difficult to design a globally good and robust control structure for all possible cases, especially when the system works in poorly predictable situations. One of the best solutions is the formation of a finite number of knowledge bases of a fuzzy controller for a variety of fixed control situations. The goal of a quantum regulator is to combine the knowledge bases obtained with the help of the soft computing optimizer knowledge base into self-organizing quantum fuzzy regulators. The QFI model uses private individual knowledge bases of the fuzzy controller, each of which designed using SCOptKB™ and QCOptKB™ toolkits.

Box "Kansei / Kawaii / Affective engineering" (*Fig. 1*) forming the knowledge about fillings of human being operator and concentrate the attention on control goal. KBs of fuzzy controllers and cognitive controllers designed with SCOptKB$^{TM}$ toolkit applying objective information of control object response from measurement system in feedback loop and affective response and will of human being operator described with new type of computational intelligence technology. The main performance of [8] to show the description of emotion estimation in Box "Kansei / Kawaii / Affective engineering" and the introduction of physical interpretation of quantum interference in cognition as quantum models of patterns.

### 2.2. Cognitive Robust Control System based on Quantum Fuzzy Inference Algorithm in Unconventional Intelligent Robotics

Strategy of intelligent cognitive control systems based on quantum and soft computing presented. Quantum self-organization knowledge base synergetic effect extracted from intelligent fuzzy controller's imperfect knowledge bases described. That technology improved of robustness of intelligent cognitive control systems in hazard control situations described with the cognitive neuro-interface and different types of robot cooperation. Examples demonstrated the introduction of QFI - gate design as prepared programmable algorithmic solution for board embedded control systems. The possibility of neuro-interface application based on cognitive helmet with quantum fuzzy controller for driving of the vehicle is shown.

*2.2.1. Introduction: human-operator factor description in control loop based on computational intelligence toolkit.* The role of human-operator in control loop was not considered in an explicit form or it was described by transferring simplified functions. The very inclusion of human-operator in control loop is often seen as a source of emergency and increasing risk according its decision making. It has so far proved [1] that in control of big multiple control loops connected in a single management system with human-operator, more to 75% of information often are an excess quantity which either not used or interferes decision-making process. In addition, modern control objects are complex, poorly formalized dynamic systems that are characterized by informational uncertainty of model structures and control goals, a high degree of freedom and significant nonlinearities with cross-coupling, instability, distributed sensors and actuators, high noise levels, abrupt changes in structure and dynamics, etc.

These are typical information sources for hazard and unpredictable control situations. The design of reliable modern control systems for unforeseen control situations is the cornerstone of modern control systems theory. The degree to which the control system successfully copes with the above difficulties depends on the intelligence level of the improved control system.

Soft computing technologies, such as genetic algorithms (GA) and fuzzy neural networks (FNN) had expanded application areas of fuzzy controller (FC) by adding optimization, learning and adaptation features [23-28]. For complex and ill-defined (fuzzy) dynamic control objects that are not easily controlled by conventional control systems (such as proportional-integral-derivative – PID-controllers) – especially in the presence of fuzzy model parameters and different stochastic noises – the System of Systems Engineering methodology provides FC as one of alternative way of control systems design [29].

It is difficult to design optimal and robust intelligent control system, when its operational conditions have to evolve dramatically (aging, sensor failure and so on). Despite the fact that you can predict such conditions, it is difficult to cover such situations by a single knowledge base (KB). Using unconventional computational intelligence toolkit like a Soft Computing Optimizer (SCO) and Quantum Fuzzy Inference (QFI), we propose a solution of such kind of generalization problems by introducing a self-organization design process of robust KB-FC that supported by the QFI based on quantum soft computing ideas [30-34].

So far, the theory and design processes of intelligent control systems (ICS) as KB - control systems (in the form of a relevant KB) was carried out by experts themselves through computational intelligence as soft computing using GAs or FNN.

*Problem's formulation.* The learning and adaptation problems of FC's are the important topic in control theory. Many existing solutions are using different models of artificial neural networks based on the back-propagation (BP) algorithm, Kohonen multilayer structure and so on [30]. Unfortunately, methods based on BP-algorithms and iterative stochastic algorithms do not guarantee the required control robustness level and accuracy in complex unpredicted control situations. Such schemes are successfully working if the control task performed in absence of underdetermined stochastic noises in world environments, in sensors, in control loop, etc. Therefore, one of the central problems of developing cognitive control system (CCS) was in the finding a constructive solution of design KB tasks and intelligent robust cognitive control in a given problem-oriented application. On the other hand, one of the key tasks of modern robotics is the development of technologies of cognitive mechanical interaction, which allow solving intelligent control functions due to the redistribution of knowledge and control on the software level. The solution of this task based on Soft Computing Optimizer (SCO) is developed in [29]. QFI model including allows self-organization level in intelligent control system [31-34]. QFI applying the laws of quantum computing technologies [35] and applies next operations: superposition, quantum correlations (entanglement or quantum oracle), and interference [35-37].

*2.2.2. Method of solution and its physical background.* Proposed QFI system consists of a few FCs, each of which provides solution in one set conditions of control system. QFI system revises the results of fuzzy inference of each independent FC and proposes in on-line the generalized control signal output. The output QFI signal combines best features of each independent FCs. Extracted quantum hidden information amount from classical control states considered as additional powerful recourse of thermodynamic entropy control force and quantum intelligent force control realized on useful work of cognitive controller.

New approach for robust cognitive controller design applied based on quantum information thermodynamic law [9]. The changes of entropy and the quantum mutual information lay new limit for the marginal part of work which exceeds the conventional second law of thermodynamics. Surprisingly, if two measurements are preformed, it find a new inequality due to the entropic uncertainty relation with the assistance of quantum memory, which provides a lower bound for the work gained from the heat engine. This result describes an opposite fact of the thermodynamics by considering entanglement in quantum information science [10].

Quantum soft computing optimizer toolkit of KB - design processes based on QFI model is described. Benchmarks of robust KB design from imperfect FC - KB as the new quantum synergetic information effects of extracted quantum information demonstrated. Moreover, the new force control law from quantum thermodynamic described: with extracted hidden quantum information from classical control signal states (on micro-level) possible to design in on-line new control thermodynamic force that can produce on macro-level more value work amount than the work losses on the extraction of this amount of hidden quantum information [8-10]. It is a new control law of physics-cybernetics open hybrid systems including port-Hamiltonian controlled dynamic objects [10] and as method design of robust cognitive controller applied.

*Main goal.* The main purpose of QFI is to produce a self-organization capability for many unpredicted control situations. QFI produces robust optimal control signal for the actual control situation using a redundant amount of information in KB's of individual FCs [8]. In this work the main ideas of quantum computation and quantum information theory [18,35-37] applied in developed QFI methods are introduced. Robustness of new types of self-organized intelligent control systems is demonstrated [10].

*2.2.3. Structure of QFI based on intelligent FC and cognitive controller.* QFI design is needed to apply the additional operations as superposition, entanglement (quantum correlation or quantum oracle) and interference that physically are operators of quantum computing. Let us introduce briefly the particularities of quantum computing and quantum information theory that are used in the quantum block QFI (see *Fig. 10*).

QFI support a self-organization capability of FC in robust intelligent control system (ICS). In *Fig. 10* FC1 is intelligent fuzzy controller and FC2 is cognitive controller [27,28].

*Quantum computing.* The fundamental result of quantum computation stays that all of the computation can be embedded in a circuit, which nodes are the universal gates. These gates offer an expansion of unitary operator *U* that evolves the system in order to perform some computation. Thus, there is two problems:

(i) Given a set of functional points $S = \{(x, y)\}$ find the operator U such that $y = U \cdot x$;

(ii) Given a problem, fined the quantum circuit that solves it.

Algorithms for solving problems of quantum computing may be implemented in a hardware quantum gate or in software as computer programs running on a classical computer. It shows that in quantum computing the construction of a universal quantum simulator based on classical effective simulation is possible in [8,18,34,35].

In the general form, the model of quantum algorithm computing comprises the following stages:
- preparation of the initial state $|\psi_{in}\rangle$;
- application the Hadamard transform for the initial state in order to prepare the superposition state;
- application of the quantum correlation operator to the designed superposition state;
- application of the interference operator;
- application of the measurement operator to the result of quantum computing $|\psi_{out}\rangle$.

Hence, a quantum gate approach can be used in a global optimization of KB structures of ICSs that are based on quantum computing, on a quantum genetic search and quantum learning algorithms. *Figure 11* describes the functional diagram and structure of QFI.

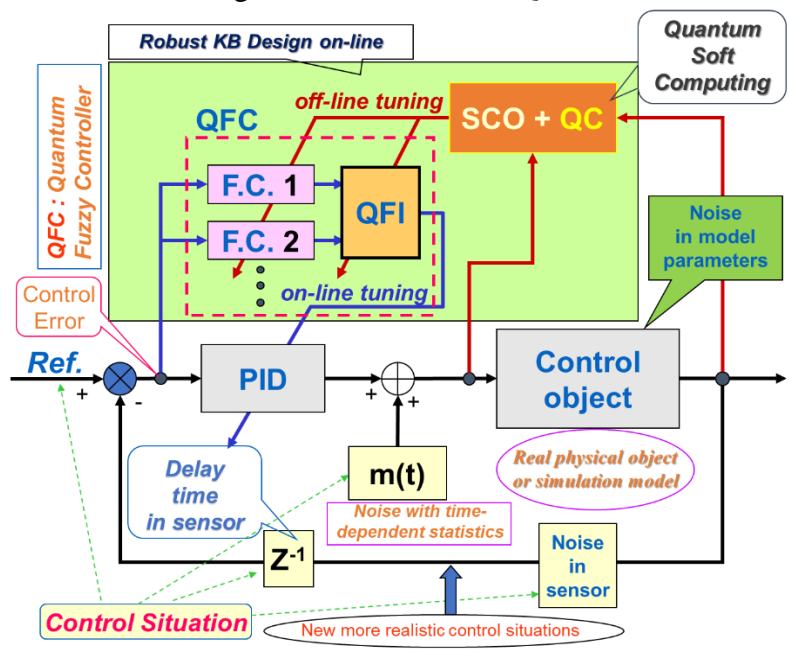

*Figure 10: Structure of robust ICS based on QFI model.*

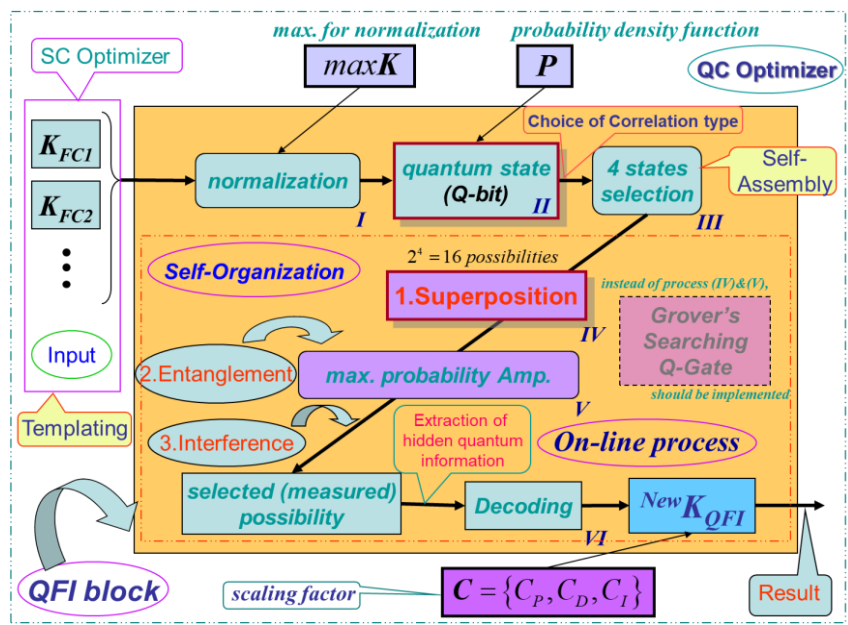

*Figure 11: The functionality structure of QFI.*

*Figure 12* describes programmable algorithmic gate of QFI realization of which can be implemented using both classical and quantum computer, and can also be integrated into different control system and embedded intelligent controllers.

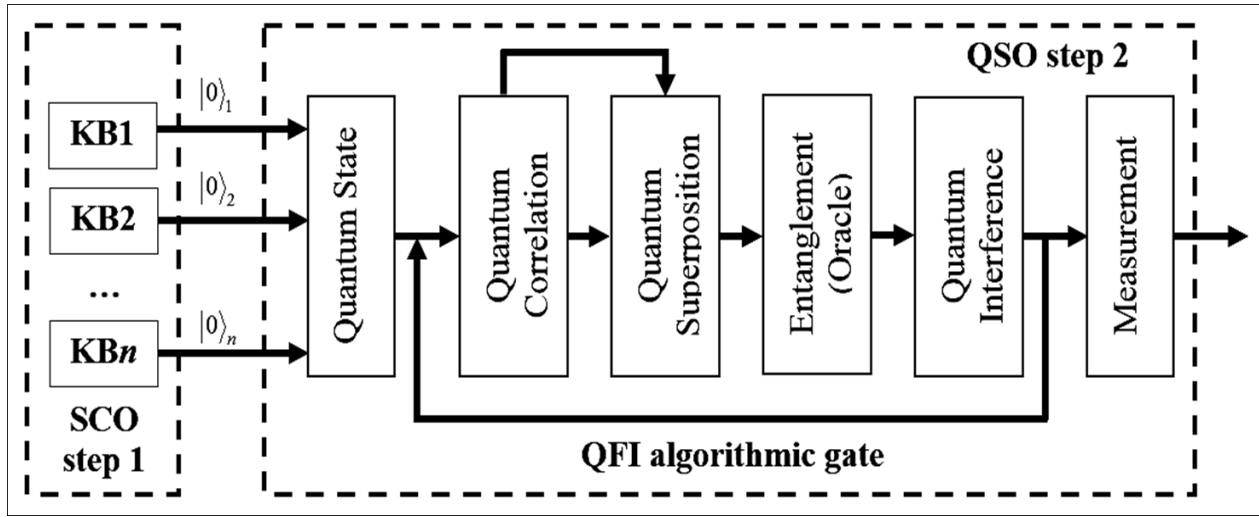

*Figure 12: Quantum algorithmic gate of QFI realization.*

This algorithm is based on quantum information theory as following: superposition, quantum correlation (entangled operators), and interference. The fourth operator, measurement of result quantum computation, is classical and irreversible. Optimal control process based on value information amount that extracted from a few KBs (that are designed by soft computing) and are founded on the following facts from quantum information theory:

(i) the effective quantum data compression;
(ii) the splitting of classical and quantum parts of information in quantum state;
(iii) the total correlations in quantum state are "mixture" of classical and quantum correlations;
(iv) the exiting of hidden (locking) classical correlation in quantum state.

If these facts are used, the extraction an additional amount of quantum value information from KBs becomes possible to wise control organization. These facts are the informational sources of QFI background. That is QFI using compression and rejection procedures of the redundant information in a classical control signal.

## 3. Remote quantum knowledge base optimization

Scaling factor is used in remote quantum knowledge base optimization as the adjustable parameter. Scaling factor is used in the final step of forming the gain of PID.

During operation, data in symbolic form are passed via COM-port. The control system reads data from the sensors and sends them to a computer for further processing. By taking the input values, the GA evaluates the previous decision, and carries a QFI to check the following solutions. The result of the QFI is sent to the remote device for generates of control action.

### 3.1. Benchmark: A neuro-interface based cognitive control

In order to increase the effectiveness of intelligent management and the guaranteed achievement of the management goal in conditions of uncertainty, contingencies and increasing information risk, the idea of applying cognitive processes occurring in the cerebral cortex of a human operator has emerged. In this case, the processing of cognitive processes, in the form of electric signals of electroencephalograms (EEG) and recorded by the brain-device interface from the cerebral cortex parts, allows designing additional KBs based on the KB optimizer on soft computing and forming additional control forces to achieve the goal control.

As a result, such a combination of the cognitive regulator with the technical intelligent controllers in the control loop leads to the formation of a new class of control systems - cognitive intelligent control systems, which themselves can be considered as a unified socio-technical system. KBs of cognitive and intelligent controllers are designed and created based on a control error signal and corresponding training signals that describe the dynamic behavior of the control object or EEG of the human brain. The control error contains data about an unforeseen control situation, and also acts as input data for the considered controllers. The output signals of the regulators reflect the reactions of the KB, which in the general case are inaccurate (imperfect) KBs and can lead to a loss of robustness of the control system. Both signals are inputs to the QFI block, which performs self-organization of incomplete KBs, forming in real time a new robust KB of a hybrid quantum intelligent cognitive controller.

Currently, for the training of cognitive abilities of the operator are widely used simulators games. One of the main problems of the application of the cognitive system control technologies "brain - computer" is the problem with the physiological features of the human being operator. Exhaustion, excitement, distracting noises, etc. affect the physical condition that naturally affects recognition quality teams and quality of control target device. Now, do not possess specialized simulators software module for adaptation and learning control system (the program itself) to the characteristics of the operator [8]. Experience with such weights indicates that virtually no mechanisms of adaptation and learning for practices imposed by the entry signal and its repetition, which interpreted as the generation of commands. For the experiment was been select the object of control-mobile robot in the form of three-wheel vehicle with Bluetooth-control (see *Fig. 13*).

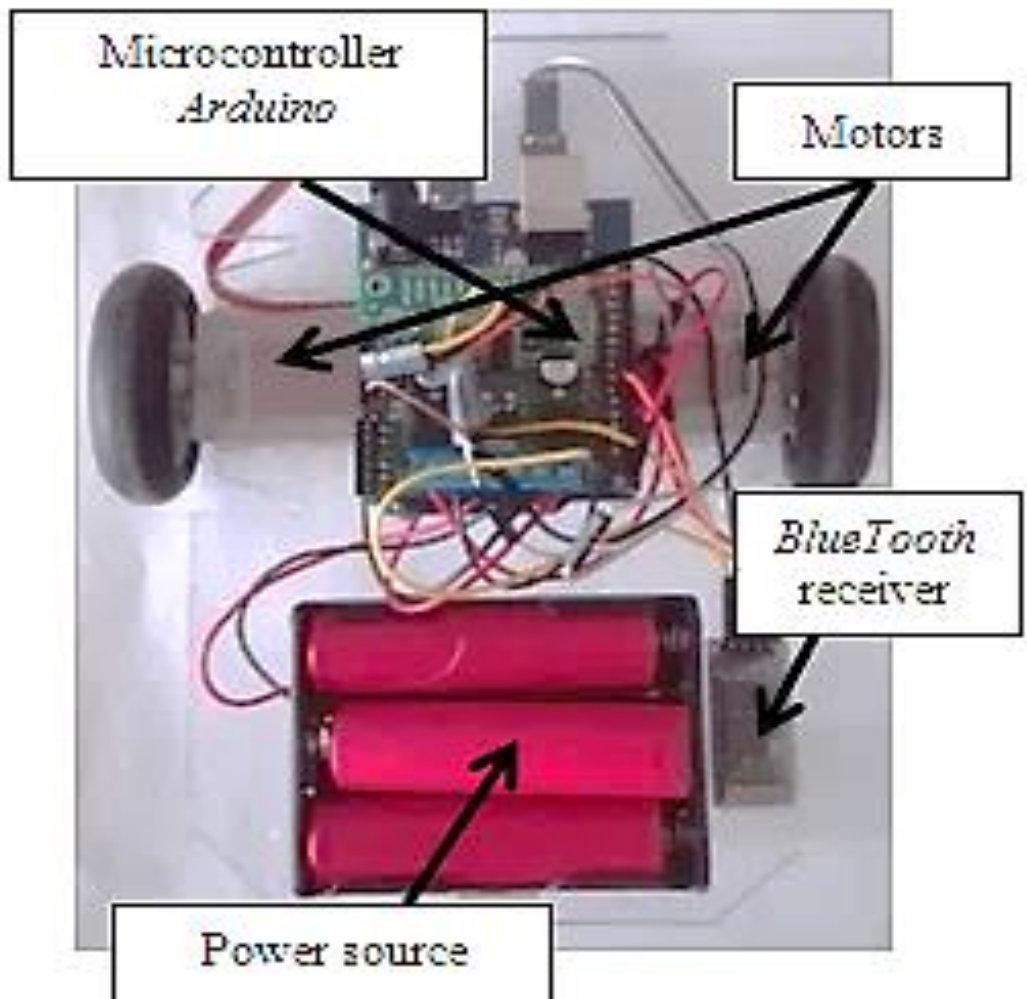

*Figure 13:* Control Unit.

The hardware and software parts:
- Robot-car based on Arduino UNO controller;
- Engine driver Pololu Dual MC33926
- 2 motors DC 9V;
- Bluetooth module HC-05;
- Power supply as 3 3.7V Li-On batteries;
- Neurointerface OpenBCI with EEG Electrode Cap Kit;
- PC on Windows 10 with Bluetooth module;
- SCOptKB software toolkit.

To approximate the teaching signal applied the developed SCO with selected the model of fuzzy inference (Sugeno type models). As teaching signal is used the signal from the block signal recognition of cognitive device [38], as well as the integral value of the signal.

Generalized technology design represented as a diagram presented in *Fig. 14*.

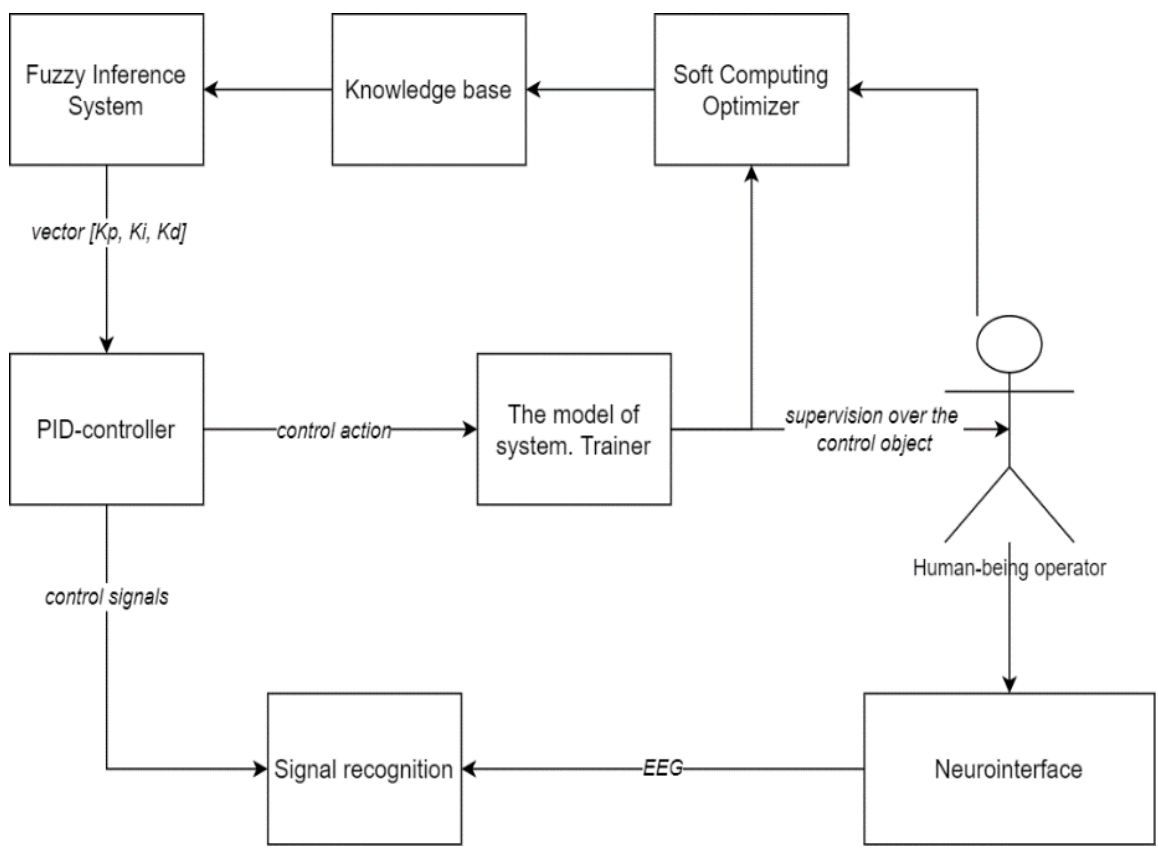

*Figure 14: The structure of the design technology of cognitive control systems with SCO.*

In such a structure control system with model is a central element, that rendering action that needs to be done. For the control of vehicle, it can be a game where you must control the machine.

***Related works.*** Quantum computing approaching in robot path planning, navigation, learning, emotion design, decision making was applied in [8-11 and references in their] etc. Our approach is based on quantum self-organization of KBs using responses of FCs on unpredicted situations in online and applying in cognitive interface.

The structure of cognitive intelligent control of autonomous robot based on IT of robust FC design and neuro HMI in *Fig. 15* demonstrated.

On *Fig. 16*, move to right corresponds to + 1, and the movement to the left corresponds to – 1.

*Figures 16 - 19* presents the results of an experiment using FC and quantum controller with QFI. Differential component in FC associated with the speed of the operator activates the mental command.

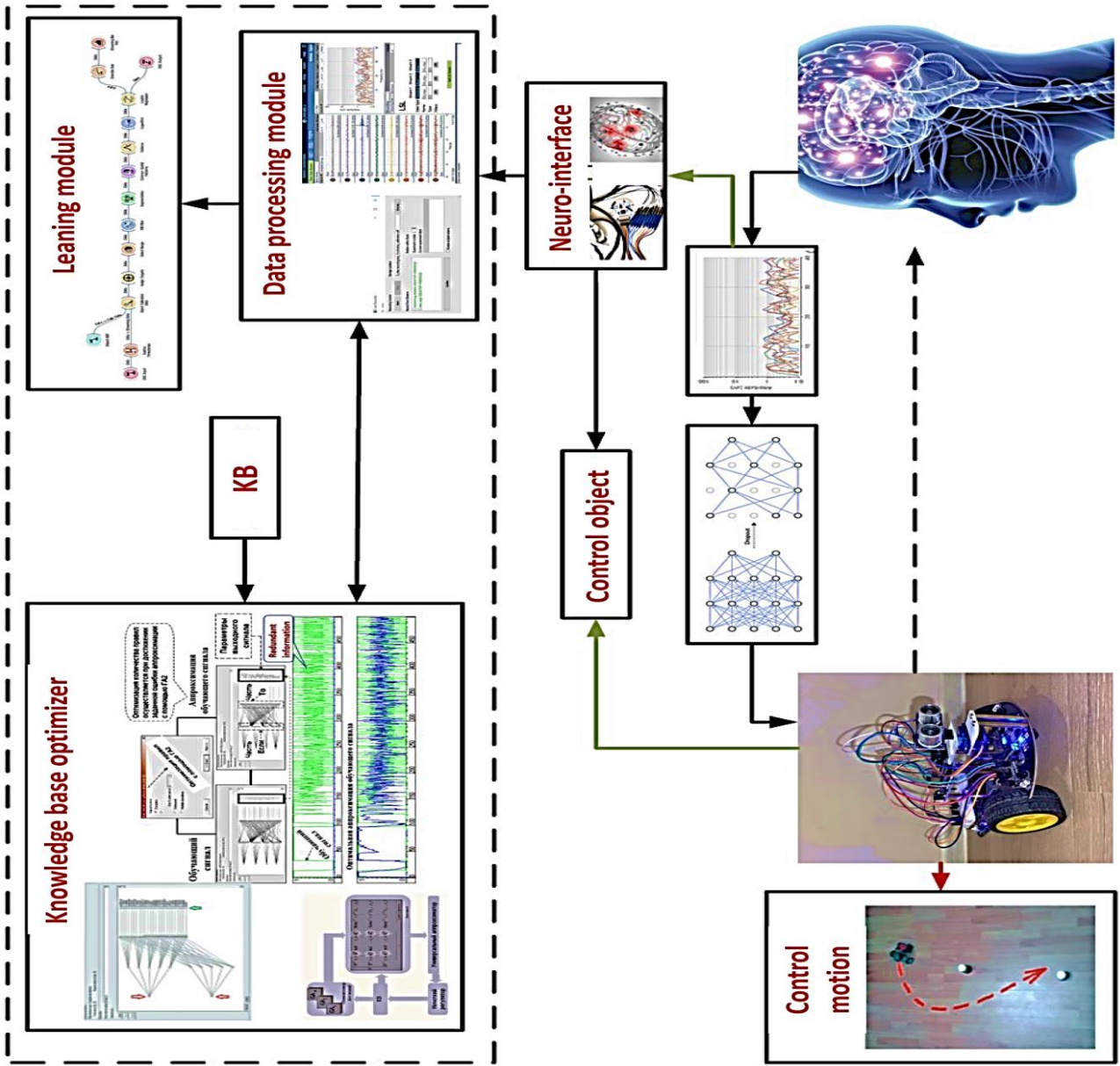

*Figure 15: The structure of cognitive intelligent control of autonomous robot based on IT of robust FC design and neuro HMI.*

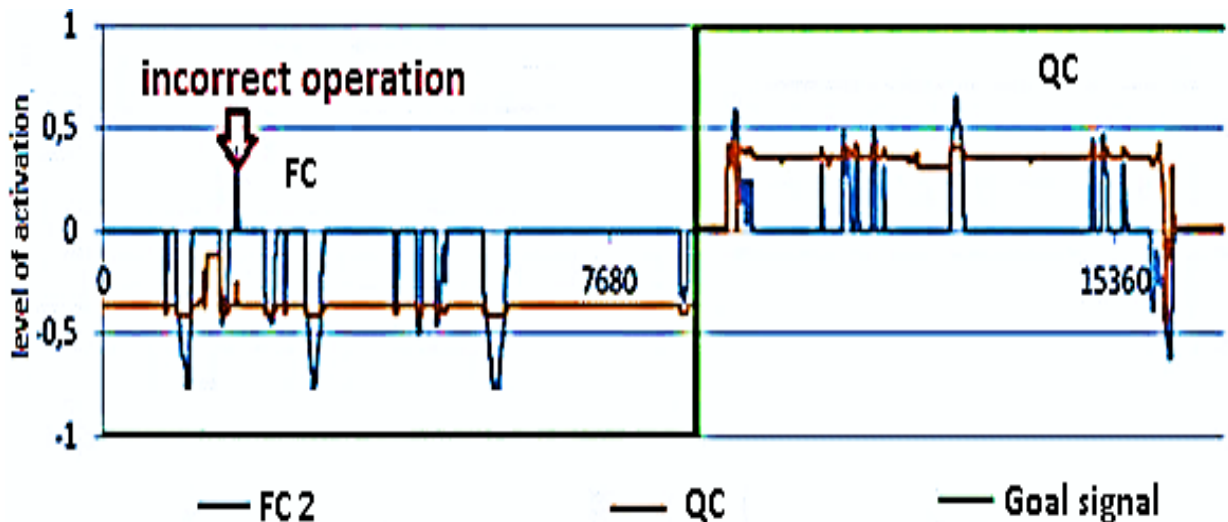

*Figure 16: Control actions produced by fuzzy controller and quantum controller adjuster when moving left and right.*

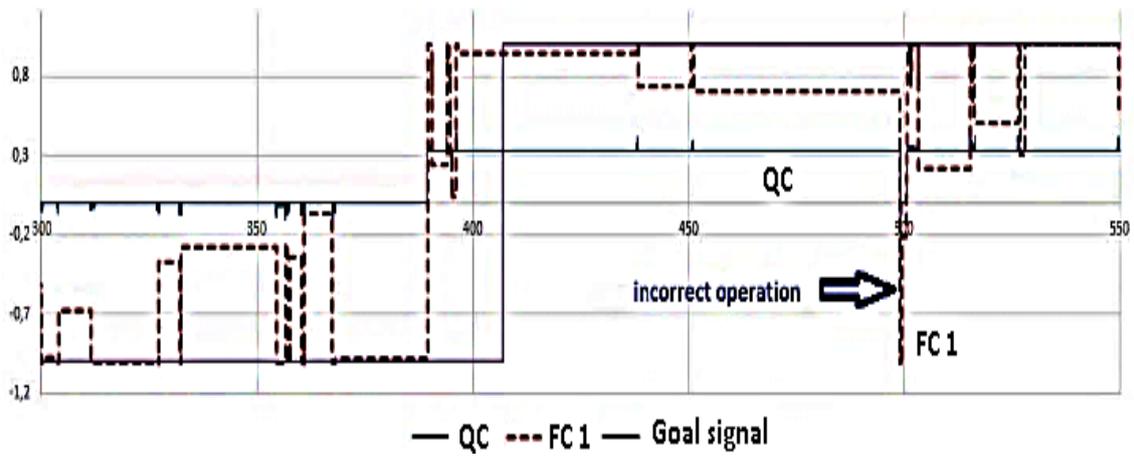

*Figure 17: Control actions produced quantum controller and fuzzy controller when driving forward and back.*

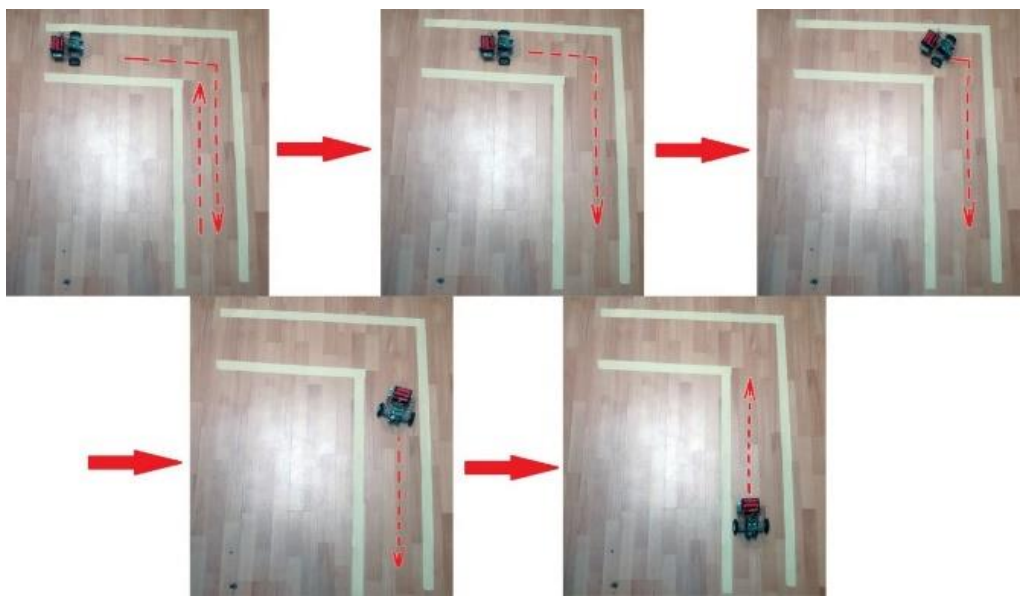

*Figure 18: When controlling the trajectory of mobile robot based on the quantum controller.*

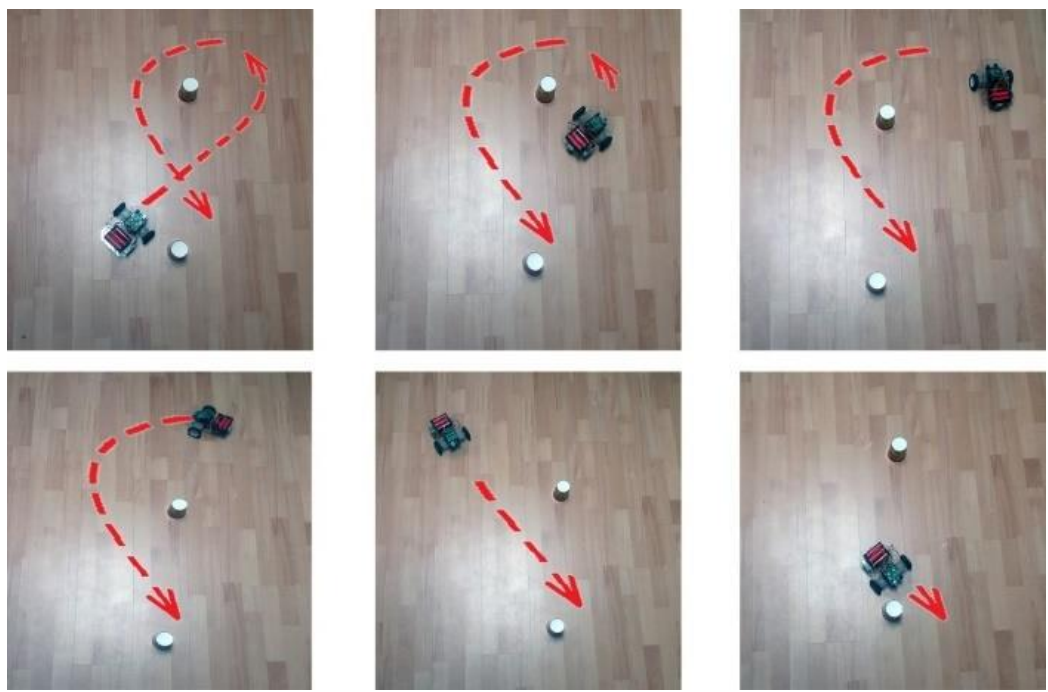

*Figure 19: Avoidance obstacles control system with quantum controller.*

The results presented on the *Fig. 17* and *Fig. 19* showed that quality of control greatly improved with intelligent cognitive control scheme based on SCO and QCO. This work of robot demonstrates the need to develop a unified information technology of design of control system for neuro-interface [8].

It is worth noting that the further development of cognitive technologies control inseparable with brain training methods. Now, operators using specialized filters, smoothing signals of the EEG and remove interferences and noises caused by psycho-physiological condition and external factors, but for global better the use of intelligent cognitive technologies, the software level of control system should be having executive mechanisms for learning, adapting and self-organization the control system.

The development of robotic human limb prostheses with non-invasive interfaces is considered as the example of the application and approbation of the developed technology. Such interfaces are actively used for rehabilitation and diagnostic procedures and help to more closely interact with the human environment, including robot for service.

### 3.2. Cognitive intelligent control of the prosthetic arm: quantum soft computing approach

The development of robotic human limbs prostheses and the production of human-like electromechanical devices - anthropomorphic robots are receiving more, but not sufficient, attention in both the scientific, technical and socio-economic plans [38-41]. It should also be noted that in the strategy for the development of artificial intelligence (AI) in the Russian Federation [42] in the «Healthcare» section, the application of AI in such an important socio-technical domain as intelligent prosthetics and smart cognitive control systems, as well as rehabilitation of disabled people, is not indicated at all.

Let us consider briefly main principles, peculiarities and features of cognitive intelligent control applied in biomechanical products and presented the description of a hierarchical intelligent control system based on QSCOptKB$^{TM}$ (knowledge base optimizer on quantum soft computing).

*3.2.1 Design Principles and Features of Biomechanical Product Quantum Cognitive Intelligent Control.* Sources of technological and breakthrough innovations in these areas are: new technologies for creating intelligent materials; technologies for creating an intelligent software product integrated into devices and applied at all stages of interaction with devices; new human-machine interfaces, the principles of which are based primarily on the method of reading the activity of the functioning of the brain and nerve endings of the body [8].

So, the *first* direction allows a person to restore (and in the future, to exceed) the functional state of limbs damaged as a result of any injuries due to the creation of more advanced alloys, material structures, nano coatings. The *second*, innovative direction is associated with the creation of sophisticated software that allows the biomechanical device to learn and adapt to individual physiological and psychological qualities and characteristics of the human operator. Through the application of deep machine learning and medical recommendation systems with a deep knowledge representation, it is possible to recognize more complex human commands, read and recognize the emotional state of the operator [6-8].

At the same time, a certain computing basis in the form of embedded end-to-end technologies of cognitive computing and computational intelligence must correspond to software of this level. The development of the *third* direction is based on the sources of new human-machine interfaces that can effectively complement and expand human information capabilities. Such interfaces include infrared - spectrometers, electroencephalographs, magnetoencephalographs, cognitive chelmets and equipment for virtual and complementary (augmented) 3D- and 4D-reality, invasive and non-invasive sensors and beacons, for example, mounted on the wrist, or other parts of the human body.

*Related works.* Many researches are known in which patients routinely apply such interfaces to solve everyday problems and control various devices. Interfaces are actively used for rehabilitation and diagnostic procedures, helping to improve interaction with the human environment, including with robotic devices [4-7,42-46]. Technologies are actively involved quantum end-to-end technologies in EEG data processing and educational processes at state levels [7, 8]. Research of this kind has been funded by states since the early 70's. There are a number of research collaborations on the creation and development of man - machine interfaces associated with all three areas [6, 7] etc. In particular, research in this area can be divided into the following groups:

- o Recognition research - development of devices for the diagnosis, modeling, simplification and reduction of threats to the interaction of the brain with the system.
- o Simulation of the brain mechanism - the use of neural network effects and the phenomena of the functioning of the brain in applied problems of information technology, for example, analysis and synthesis of information.
- o Restorative medicine - restoration of behavioral cognitive functions lost as a result of damage to the brain or body.
- o Elaboration - development of brain-computer systems in the feedback loop to accelerate and improve the functional behavior of the system [8].

The development of these researches made it possible to create new technologies for the neural interface to detect fundamental and interregional brain functions in online, as well as to develop complex mathematical algorithms to model brain activity and the resulting behavioral functions and reactions.

*Figure 20(a)* demonstrates afferent somatosensory signal that taken from the prosthetic device and is fed into the brain, from where the motor signal is sent back to the prosthetic limb. The nerve endings (located at the red circle) *Fig. 20(b)*, still present at the site of the amputation, send signals (red arrows) or the cortical reorganization (red star in the brain) generates the phantom limp pain. Other sensations that can be felt involve tingling, cramping, heat, and cold.

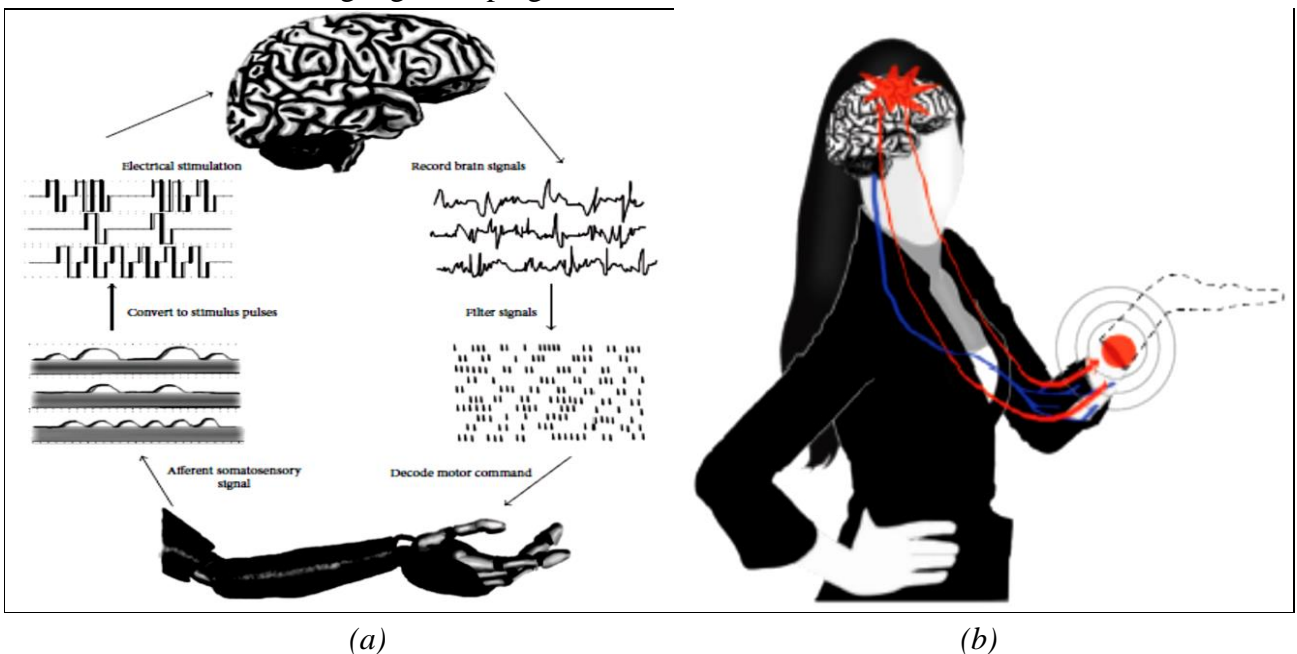

*(a)*          *(b)*

*Figure 20: Working of neural prosthetics using a brain-machine interface (a); Phantom limb pain depiction (b).*

***3.2.2 EEG signal removal and processing.*** The general concept of using a cognitive simulator is described quite fully in [8]. A well-known marker of cognitive processes is the restructuring of brain rhythms, manifested in a surface-recorded electroencephalogram (EEG) of a person. To signal the brain activity, we used the cognitive helmet company Emotiv EPOC+ (see *Fig. 21(A))*. *Figure 21(B)*, shows the structure of Emotiv EPOC + consisting of channels AF3, F7, F3, FC5, T7, P7, O1, O2, P8, T8, FC6, F4, F8, AF4 (plus CMS/DRL и P3/P4).

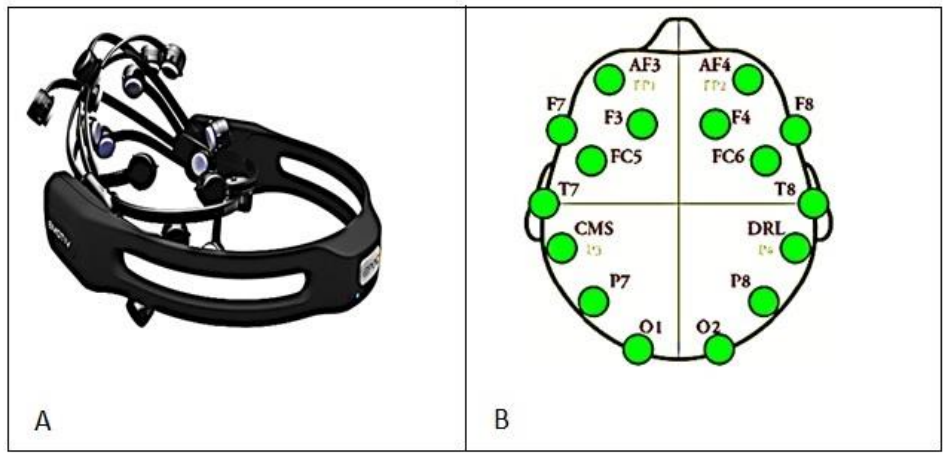

*Figure 21: (A) EmotivEPOC + cognitive helmet; (B) The layout of the electrodes of the cognitive helmet Emotiv EPOC +.*

The functional diagram of the software implementation is shown in *Fig. 22*.

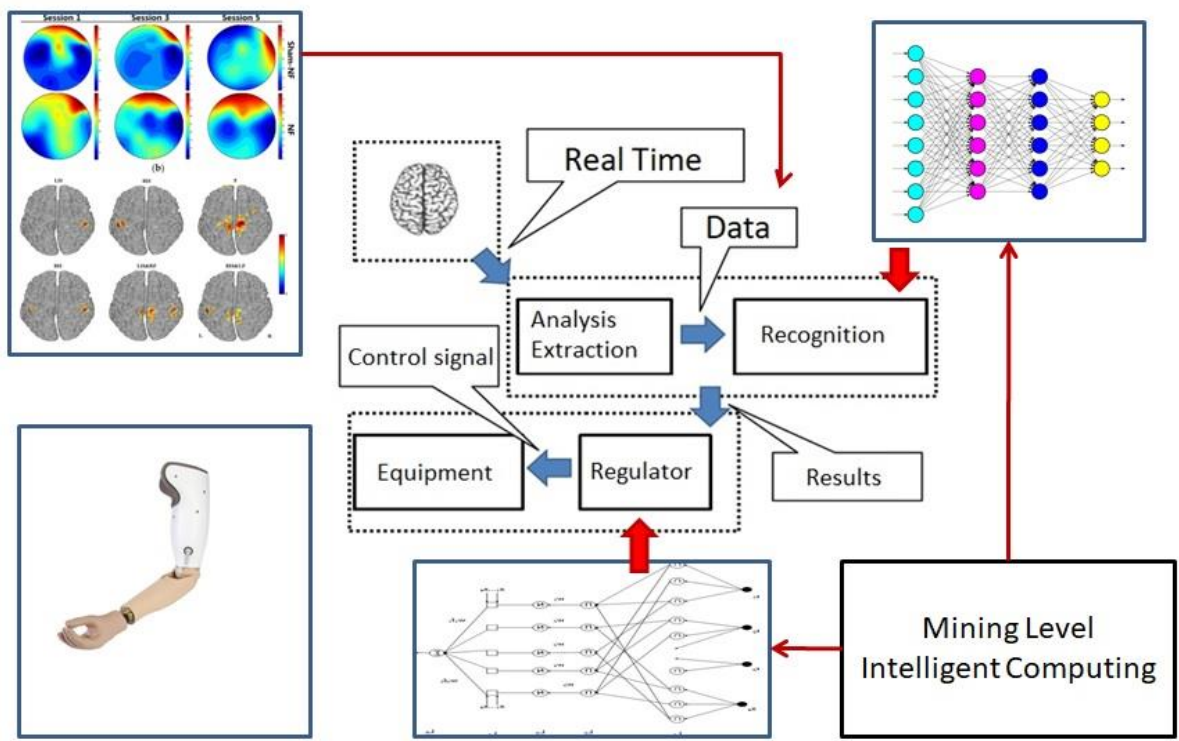

*Figure 22: Functional structure of software in on line.*

*Figure 23A* shows the EEG channel placement on the human scalp. Each scalp electrode is located at the brain centres. In 2001 Pfurtscheller (Wolpaw, 2002) identified that many of the neural activity related to fist movements are found in channels C3, C4 and Cz as shown in *Fig. 23B*. F7 is for rational activities, Fz is for intentional and motivational data, P3, P4 and Pz contain perception and differentiation, T3, T4 is for emotional processes, T5, T6 has memory functions and O1 and O2 contain visualization data. In order to remove the noise from the obtained signal, any of the suitable filtering techniques may be adopted. Further the extracted data may move for classification phase.

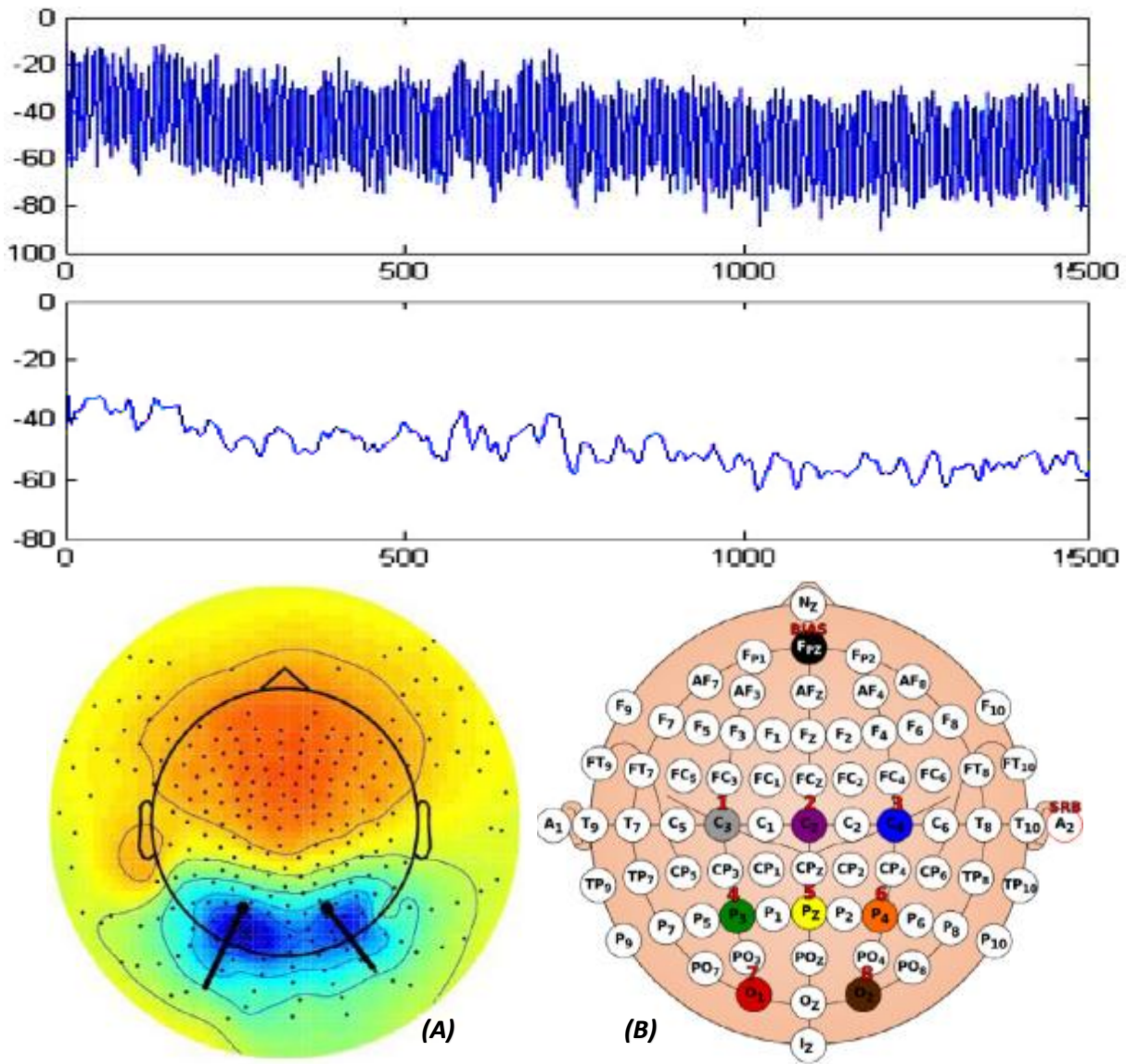

*Figure 23: EEG channel placements on the human scalp.*

In real time, the EEG data is received in the block "Analysis of the extracted data" (see *Fig. 22 - "Analysis extraction" box*). Then, after filtering and frequency decomposition, the signals enter the recognition unit. The recognition result is the degree of similarity with previously recorded commands during training. As neural networks, deep machine learning patterns of pattern recognition are used. Further, when the activation level is exceeded, such signals enter the fuzzy neural network of decision-making, designed using the soft computing optimizer. The output of such a neural network is the target values of the indicators of a managed device. At the same time, the training and operation process is supported by an emotional (positive or negative) reaction of the operator, thereby evaluating the quality of training and adaptation of the control system.

The design of the "Cognitive Regulator" block based on quantum soft computing optimizer is considered in [43]. EPOC has 14 electrodes, which are passive sensors that allow you to register electromagnetic signals. Sensors are mounted on the surface of the skin (non-submersible, non-invasive interface). The supplied software allows you to receive, recognize and record the EEG signal from the helmet. To analyze the received signal, the so-called EEG frequency rhythms are distinguished. The term "frequency rhythm" means a certain type of electrical activity corresponding to a certain state of the brain, for which the boundaries of the frequency range are defined. In the process of cognitive activity, characteristic rhythms of the beta, alpha, theta and delta ranges appear (*Fig. 24*).

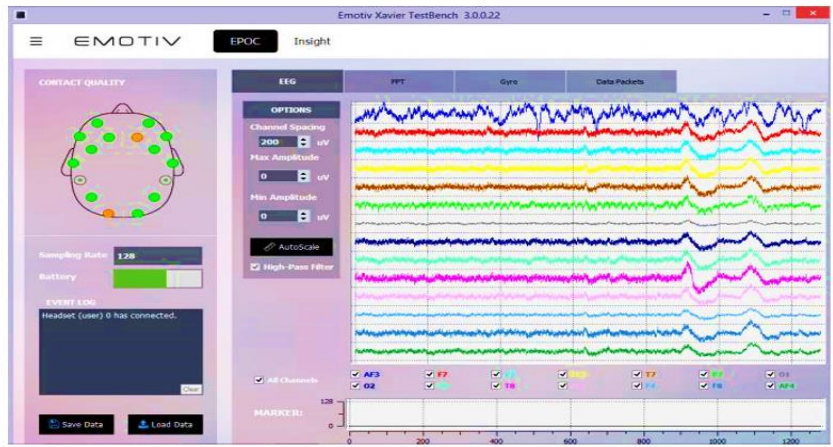

*Figure 24: EEG frequency rhythms.*

The set of simultaneously present rhythms forms a specific spatial-frequency EEG pattern. Patterns are characteristic of different types of cognitive activity and are highly individually specific. The ability of an individual to establish rhythmic EEG patterns when performing certain cognitive tasks makes up an "encephalographic" portrait of his personality.

***3.2.3 Response control of robotic prosthetics limb with BCI.*** The development of these researches made it possible to create new technologies for the neural interface to detect fundamental and interregional brain functions in on line, as well as to develop complex mathematical algorithms to model brain activity and the resulting behavioral functions and response (see, *Fig.25*).

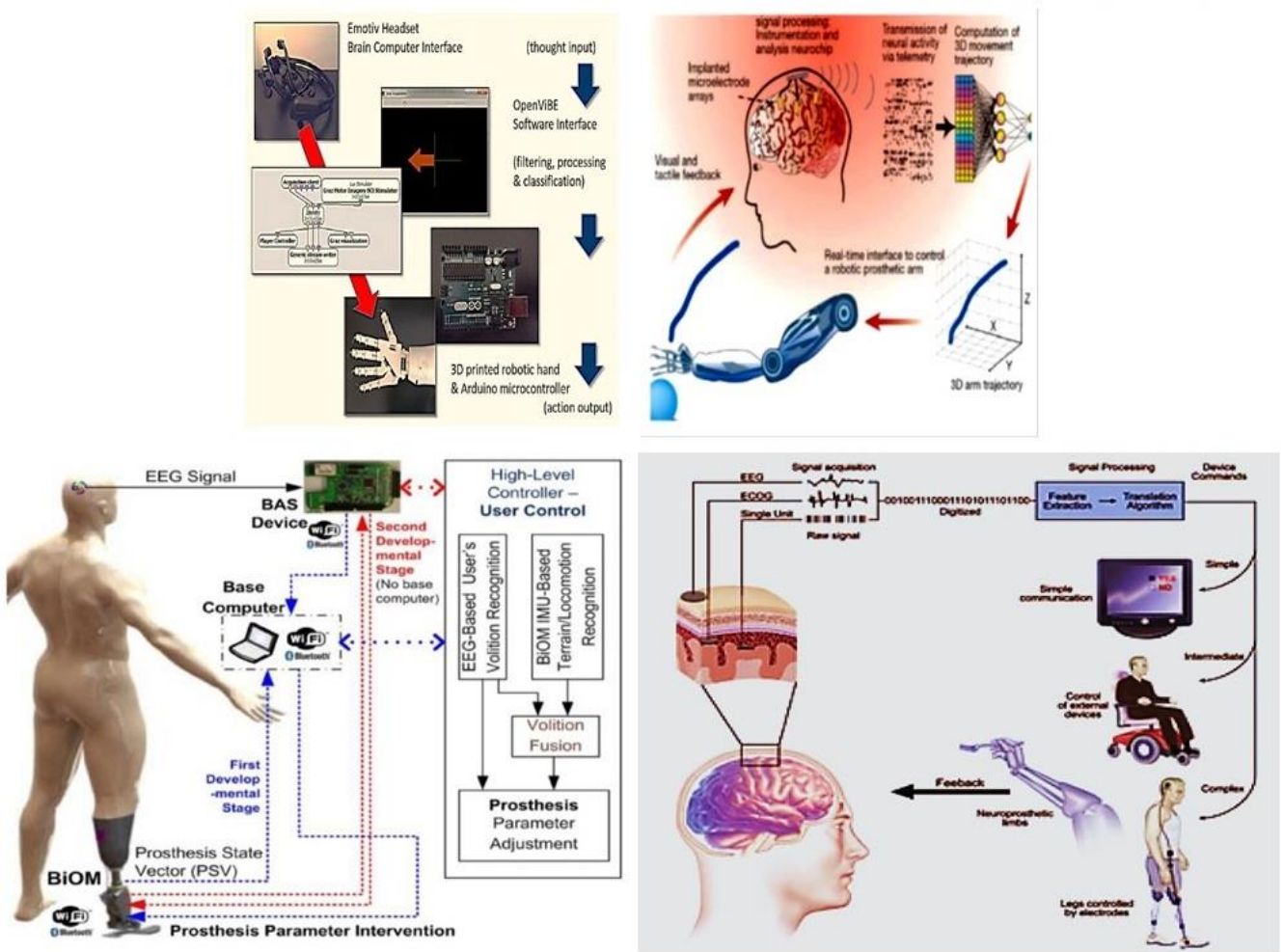

*Figure 25: Detect fundamental and interregional brain functions in on line: behavioral functions and response control of robotic prosthetics limb with BCI.*

*Figure 26* shows the generalized structure of an intelligent cognitive control system with feedback based on deep machine learning using neural networks with an optimal structure, taking into account existing approaches to the cognitive control of a robotic prosthetic arm [43].

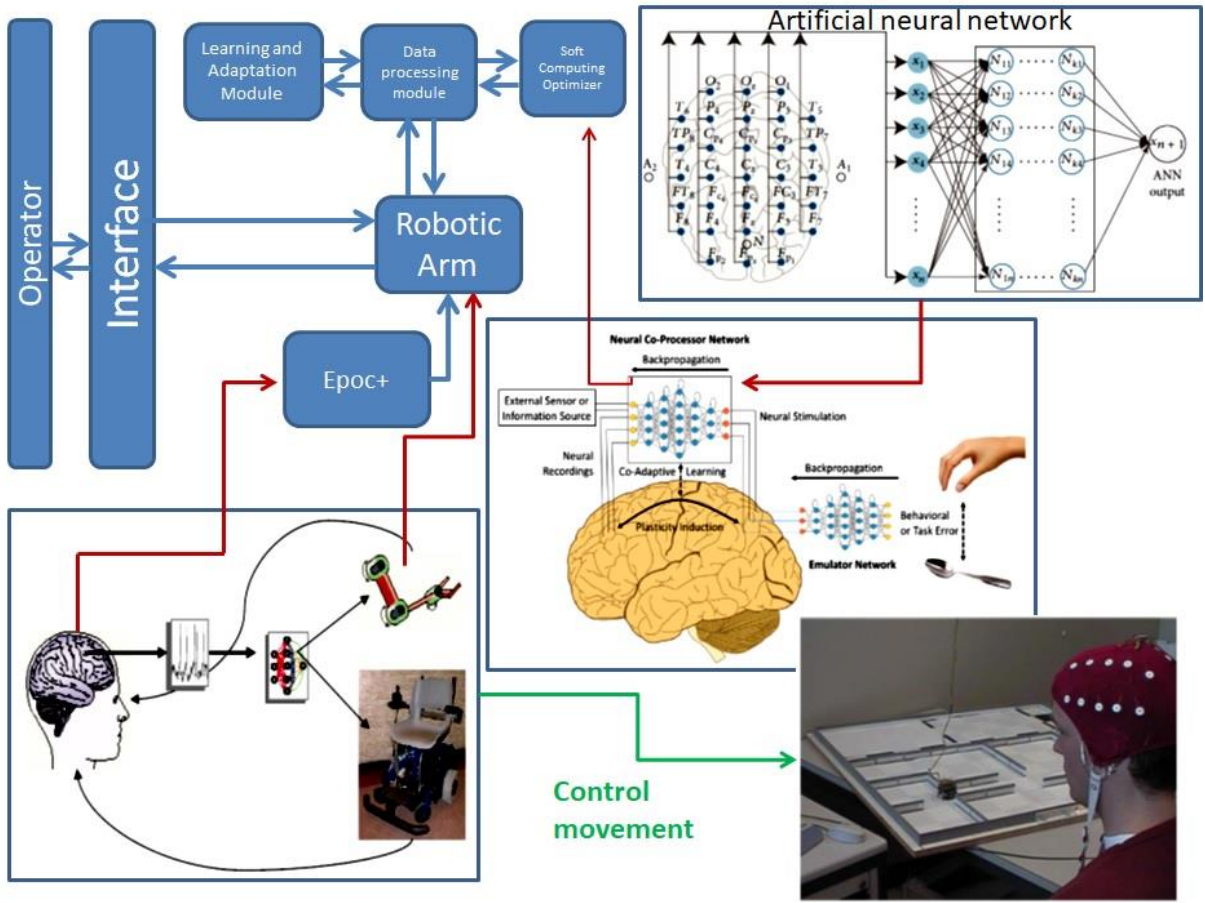

*Figure 26: The structure of cognitive intelligent control system.*

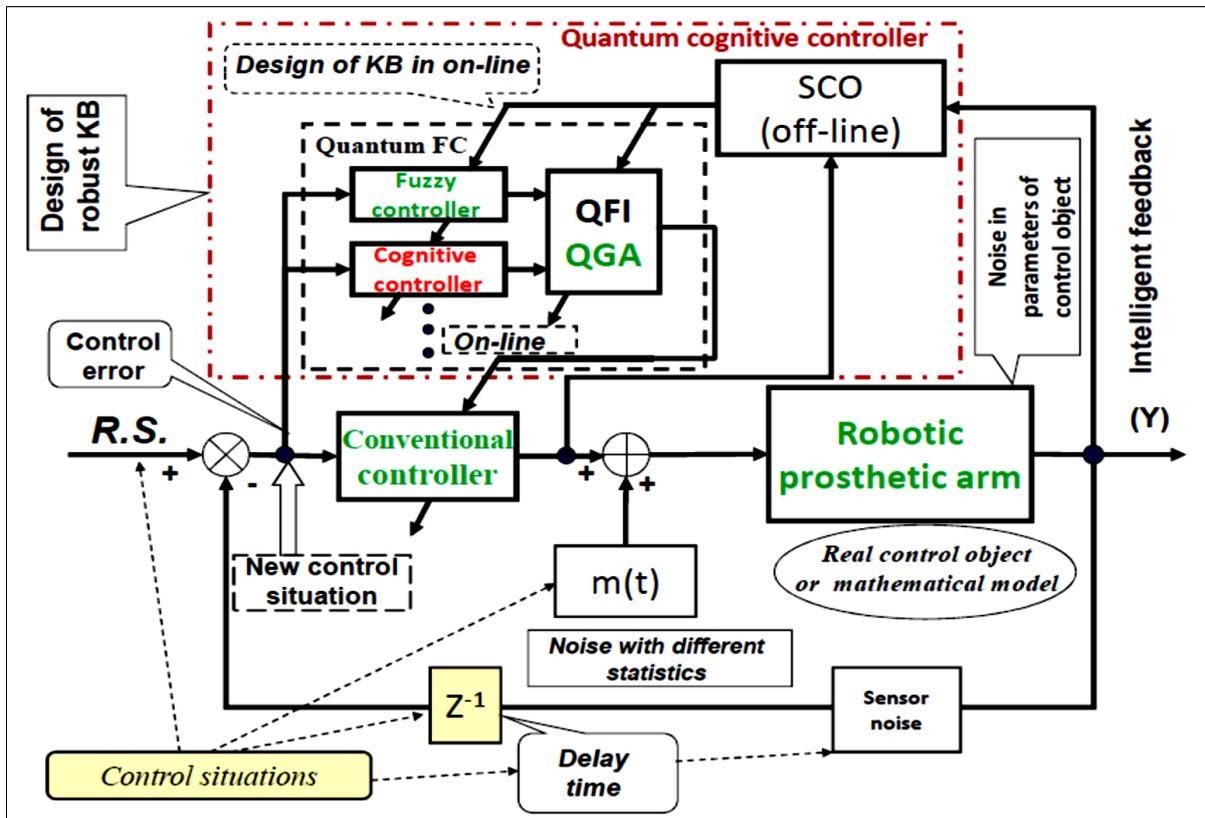

*Figure 27: Structure of Quantum cognitive intelligent control system.*

In the considered part of the work cycle, the existing hardware and software research basis and information technology of sophisticated class of cognitive intelligent control system (see, *Fig. 27* and *Fig. 1*) for supporting the design and operation of a new class of devices are presented.

Structure of cognitive intelligent control system includes two controllers: fuzzy and cognitive controllers. Design of FC and of cognitive controller KBs is achieved with SCOptKB$^{TM}$ toolkit [8-11]. In this case the responses of fuzzy and cognitive controllers in general case with imperfect KB are inputs for box "quantum fuzzy inference (QFI)" and the output of QFI is robust KB of self-organized controller for forming in on line time dependent control laws of coefficient gain schedule for traditional controller of robotic prosthetic arm.

# 4. Cognitive intelligent control of the prosthetic arm: quantum soft computing approach

Let us consider the redistribution problem of the level of responsibility between the cognitive and fuzzy controller. The basis of prostheses and robotic manipulators is spatial mechanisms with many degrees of freedom. Accordingly, when designing a cognitive intelligent controller for a prosthesis, the intelligent control system of the robotic arm can be taken as the basis. In this example, three FCs are implemented in the structure of the intelligent control system of the robot manipulator, each of which controls one of the three links independently (see in details [47]).

In *Fig. 28* shows a structural diagram of ICS by 3DOF manipulator based on KBO on soft computing with separated control.

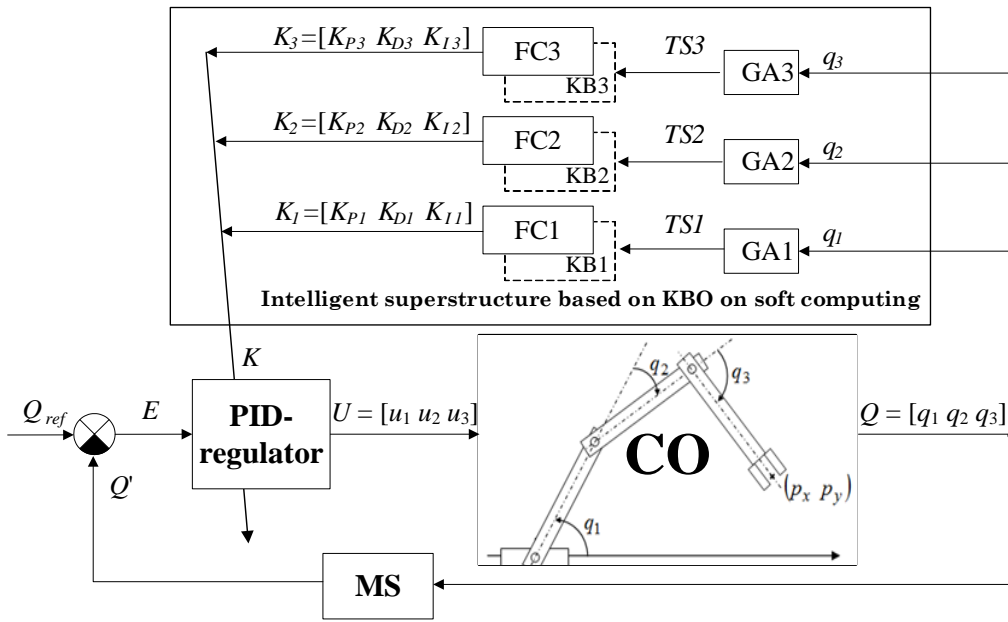

*Figure 28:* ICS with separated control based on KBO on soft computing for 3DOF manipulator

*[In Fig. 28 K means the matrix of proportional, differential and integral coefficients of the PID controller $K_{Pi}, K_{Di}, K_{Ii}, i = \overline{1,3}$, i is the number of the corresponding link of the robot of the manipulator, TSi is the training signal, GAi is a GA that generates a teaching signal for the formation of the i-th KB.].*

As a standard control situation for the FC$_i$, a typical control situation appears in the conditions of which a teaching signal *TS$_i$* received. Unexpected control situations divided into external and internal. Designed ICS based on KBO on soft computing with separated divided control (*Fig. 28*) contain information about three situations of control (standard or unexpected) for each of three links. In the designed ICS based on KBO on soft computing with separated control FC1 contains information about the standard situation 1 (KB1) for link 1, FC2 contains information about the standard situation 1 for link 2 (KB2), and FC3 contains information about the standard situation 1 for link 3 (KB3).

The scheme for extracting hidden information about the relationships between existing FCs designed using soft computing technologies for three manipulator links with KBs obtained for standard control situations using QFI unit shown in *Fig. 29*.

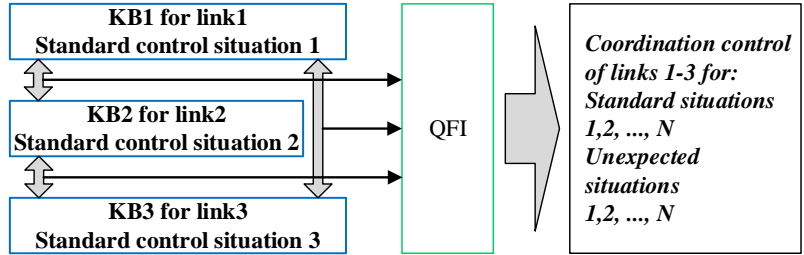

*Figure 29: Methodology for extracting hidden information of the relationships of the KB.*

We will add the QFI block to ICS based on KBO on soft computing, which can realize the self-organization of the KBs. The quantum fuzzy inference algorithm performs the following sequence of steps *Fig. 30*:

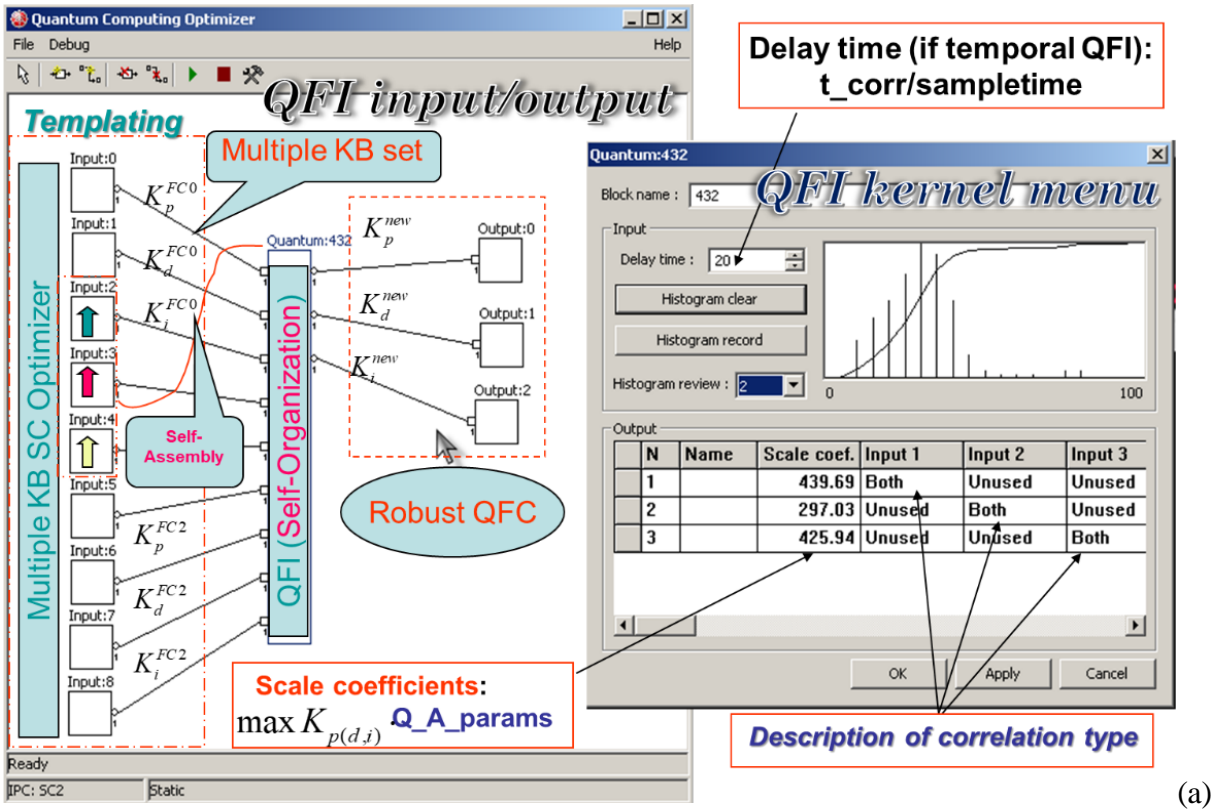
(a)

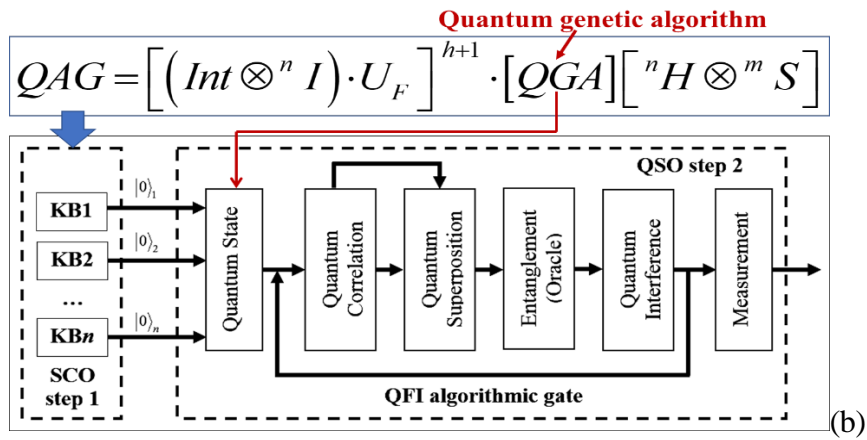
(b)

*Figure 30: Quantum fuzzy inference algorithm sequence of steps; (a) QFI kernel menu; (b) quantum algorithm gate (QAG) – approach to design of QFI based on quantum genetic algorithm (QGA).*

Steps for KB - design of quantum fuzzy controller is following:

*Step 1.* Coding.

*Step 2.* Selection of the type of quantum correlation for constructing control output signals. Three types of correlations are considered (all three types are mixed [i, ii])

*Spatial.* Output dependency $K_P^{i\_new}(t), K_D^{i\_new}(t), K_I^{i\_new}(t)$ is determined by the correlation of the following sets of input coefficients:

$$\{K_P^1(t), K_P^2(t), K_P^3(t), K_D^1(t), K_D^2(t), K_D^3(t)\} \to K_P^{new}(t)$$
$$\{K_D^1(t), K_D^2(t), K_D^3(t), K_I^1(t), K_I^2(t), K_I^3(t)\} \to K_D^{new}(t),$$
$$\{K_I^1(t), K_I^2(t), K_I^3(t), K_P^1(t), K_P^2(t), K_P^3(t)\} \to K_I^{new}(t)$$

where each set is an entangled state:
$$|a_1 a_2 a_3 a_4 a_5 a_6\rangle = |K_P^1(t), K_P^2(t), K_P^3(t), K_D^1(t), K_D^2(t), K_D^3(t)\rangle.$$

*Spatio-temporal*:

$$\{K_P^1(t), K_D^1(t-\Delta t), K_P^2(t), K_D^2(t-\Delta t), K_P^3(t), K_D^3(t-\Delta t)\} \to K_P^{new}(t)$$
$$\{K_D^1(t), K_I^1(t-\Delta t), K_D^2(t), K_I^2(t-\Delta t), K_D^3(t), K_I^3(t-\Delta t)\} \to K_D^{new}(t).$$
$$\{K_I^1(t), K_P^1(t-\Delta t), K_I^2(t), K_P^2(t-\Delta t), K_I^3(t), K_P^3(t-\Delta t)\} \to K_I^{new}(t)$$

*Temporary*:

$$\{K_P^1(t), K_P^2(t), K_P^3(t), K_P^1(t-\Delta t), K_P^2(t-\Delta t), K_P^3(t-\Delta t)\} \to K_P^{new}(t)$$
$$\{K_D^1(t), K_D^2(t), K_D^3(t), K_D^1(t-\Delta t), K_D^2(t-\Delta t), K_D^3(t-\Delta t)\} \to K_D^{new}(t).$$
$$\{K_I^1(t), K_I^2(t), K_I^3(t), K_I^1(t-\Delta t), K_I^2(t-\Delta t), K_I^3(t-\Delta t)\} \to K_I^{new}(t)$$

*Step 3.* Building a superposition of entangled states.

*Step 4.* Intelligent quantum state measurement.

*Step 5.* Decoding.

*Step 6.* Denormalization.

The overall assessment of the quality of control is higher in the case of applying ICSs on the SCO on quantum computing (for all considered types of correlations) compared to ICSs on the SCO on soft computing with divided control, which is a consequence of introducing into the structure of ICSs an additional QFI link organizing coordination management.

Moreover, if for a robot manipulator with 3DoF as a result of testing MatLab / Simulink models, it had the best performance ICSs, using spatio-temporal correlation, then physical testing determined the use of spatial correlation to be the most optimal. For a robot manipulator with 7DoF, in most cases (five out of six), spatial correlation was also optimal.

For a robotic arm of a robot with three degrees of freedom, the overall assessment of control quality is improved when using intelligent control systems on SCO in quantum computing compared to using ICSs on SCO on soft computing with one FC. ICSs on SCO on soft computing with one FC, it is able to solve the positioning problem in the conditions of external unforeseen situations, but it does not always cope with the occurrence of internal unforeseen situations.

### 4.1 Example.

In this section, 3DoF manipulator control systems considered both at the simulation level and at the physical test benchmark.

*4.1.1. Management task.* Three FCs implemented in the selected configuration of the ICS structure; each FC independently controls one of the three links.

In *Fig. 28* shows a structural diagram of ICS by 3DOF manipulator based on KBO on soft computing with separated control.

We will add the QFI block to ICS based on KBO on soft computing, which can realize the self-organization of the KBs.

*4.1.2. Modeling and experiment: control quality.* The performance of the considered ICSs evaluated based on the results of MatLab / Simulink modeling and the results of a series of experiments on CO Testbench.

Let us compare the work of ICS based on KBO on quantum computing using spatial, spatio-temporal and temporal correlations and ICS based on soft computing with separated control.

In *Figs 31* and *32*, a comparison of ICS is given for MatLab/Simulink models and manipulator Testbench according to the introduced system of quality criteria.

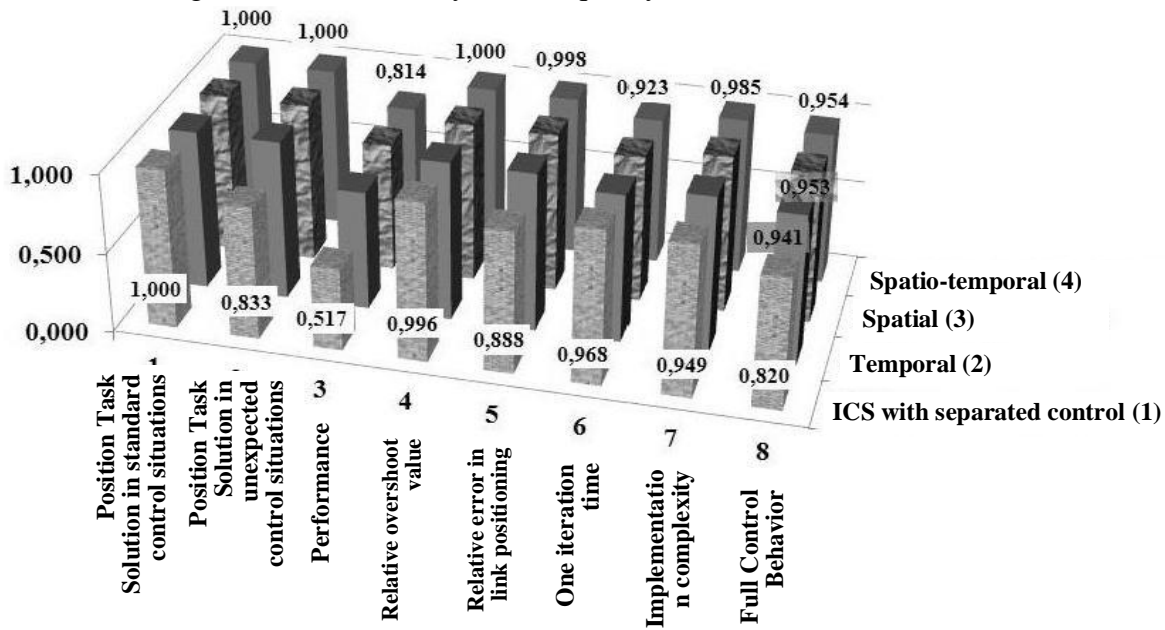

*Figure 31: Comparison of the results of ICS based on KBO on soft computing with separate control (1) and ICS based on quantum computing using temporal (2), spatial (3) and spatio-temporal (4) correlations when testing Matlab/Simulink models.*

From the comparison results, we conclude that the addition of the QFI unit to ICS with separated control the control task is solved for both standard and unexpected control situations; the performance increases; the accuracy of link positioning improves; and the implementation complexity of control decreases. The complexity of the implementation control depends on the dynamics of the control signal; we consider the robustness of the generated control laws below.

Because of QFI block apply in ICS, all quality criteria are improved as a result of eliminating the mismatch of the work of separated independent KBs by organizing coordination control. Moreover, if for MatLab/Simulink models the best indicator of full control behavior provided when using spatio-temporal correlation, and then manipulator Testbench determines the optimal choice of spatial correlation.

Next, we will consider the ICS based on KBO on quantum computing only with the use of spatial correlation.

In *Fig. 32* compares the operation of ICS based on KBO on soft computing with separated control in the conditions of an external unexpected control situation in comparison with the ICS based on quantum computing.

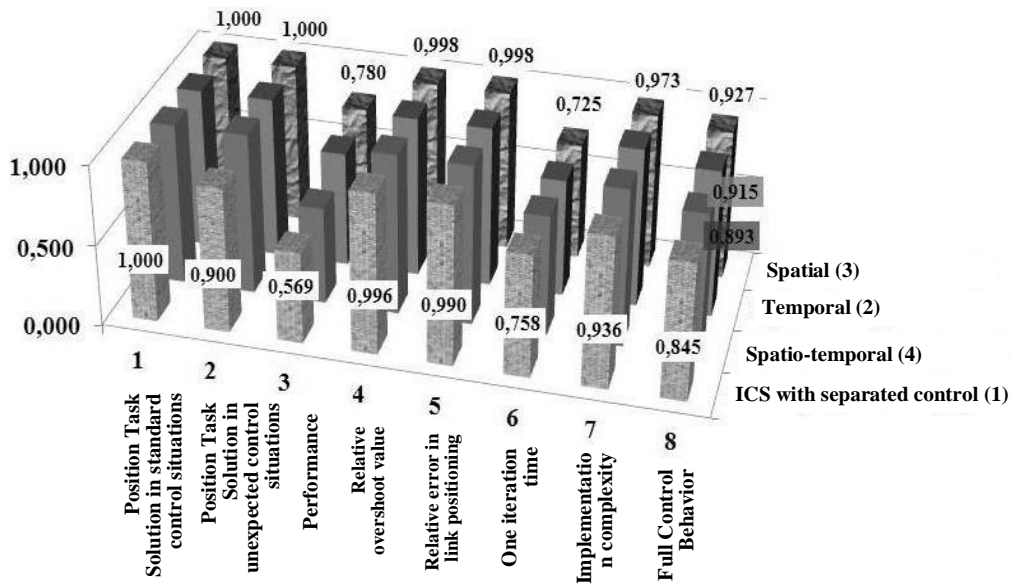

*Figure 32: Comparison of the results of ICS based on KBO on soft computing with separate control (1) and ICS based on KBO on quantum computing using temporal (2), spatial (3) and spatio-temporal (4) correlations when testing manipulator Testbench.*

The forced displacement of the second link is an unexpected situation in this case.

From *Fig. 33* it can see that in the considered unexpected control situation, ICS based on KBO on quantum computing (QCOptKB$^{TM}$ toolkit) decides with the positioning problem with a given accuracy, in contrast to ICS based on KBO on soft computing with separated control.

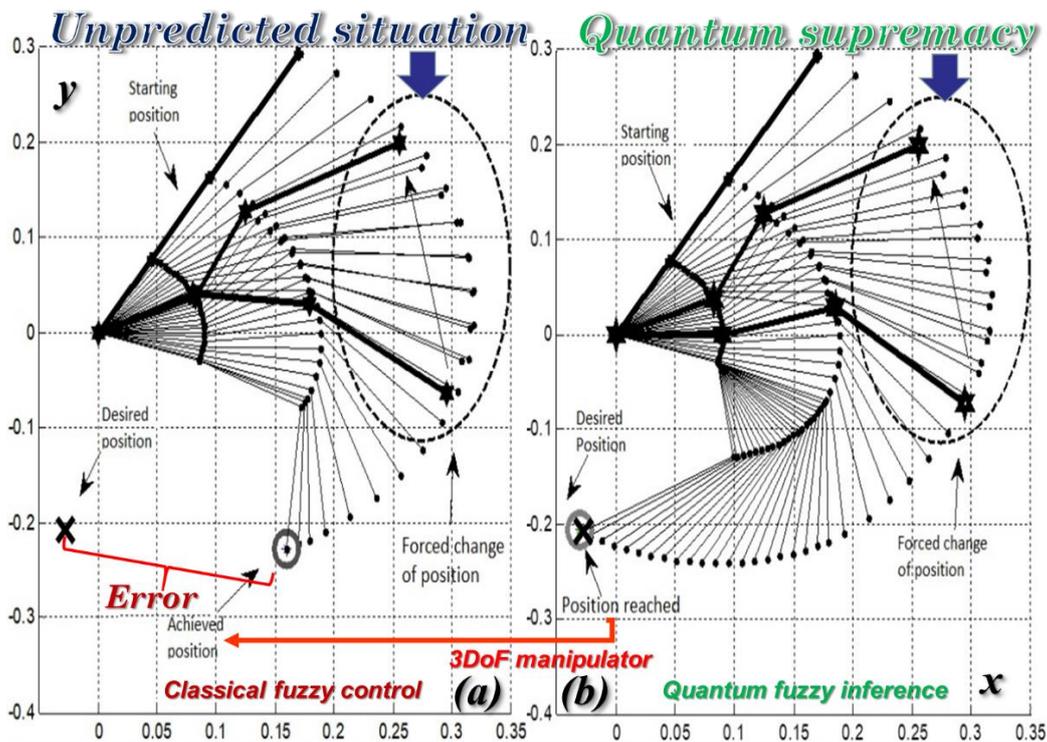

*Figure 33: ICS based on KBO on soft computing with separated control in unexpected control situation (a); ICS based on KBO on quantum computing (b).*

The inability of ICS based on soft computing to solve the problem of exact position control in *Fig. 34* also illustrated. FC responsible for managing the second link for the allotted time was not able to rehabilitate after a powerful external impact. Because of which the positioning error of the second link was more than 50 degrees, the control goal has not achieved and the control system as a whole was not robust.

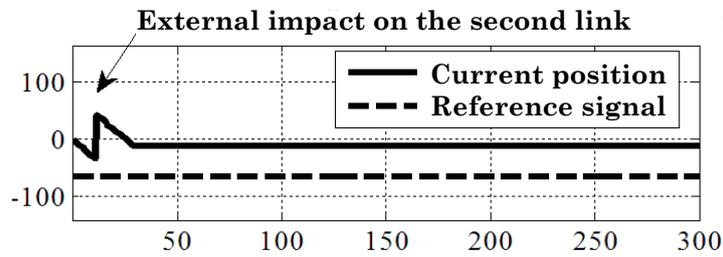

*Figure 34: The change in the position of the second link under the control of ICS based on KBO on soft computing with separated control.*

Consider internal unexpected control situations. Let us compare ICS based on KBO on soft computing with one FC and ICS based on KBO on quantum computing in the conditions of changing restrictions of the control channel.

In *Figs 35* and *36* ICS is compared respectively for MatLab / Simulink models and for manipulator Testbench, taking into account internal and external unexpected control situations.

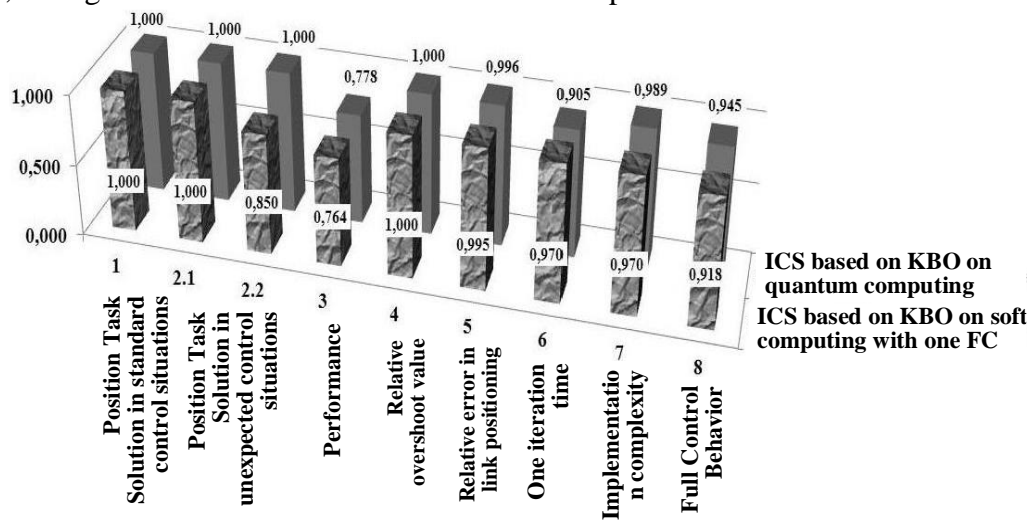

*Figure 35: Comparison of the results of ICS based on KBO on soft computing with one FC and ICS based on quantum computing when testing Matlab / Simulink models.*

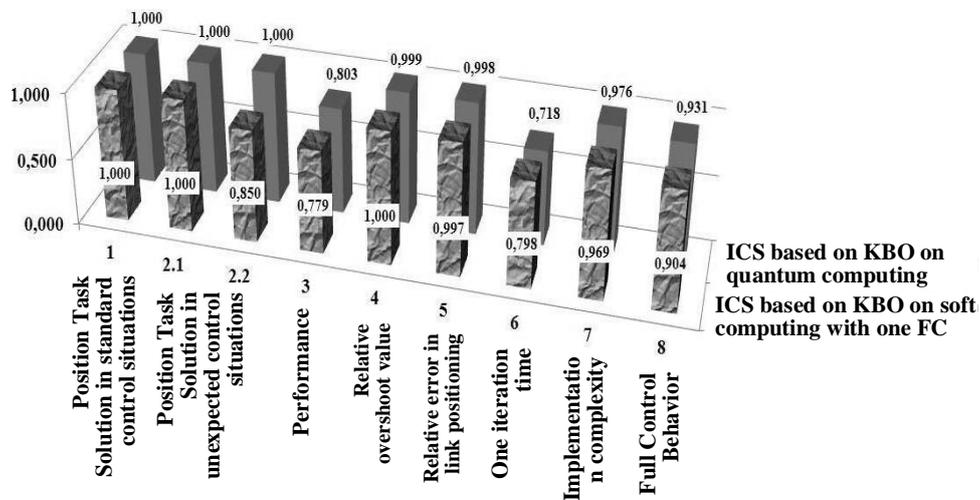

*Figure 36: Comparison of the results of ICS based on KBO on soft computing with one FC and ICS based on quantum computing when testing manipulator Testbench.*

It can see from the comparison results that ICS based on KBO on soft computing and ICS based on quantum computing solves the positioning problems for standard and external unexpected situations. ICS based on soft computing with one FC not always coping with internal unexpected control situations. Full Control Behavior for ICS on quantum computing is higher, both for MatLab/Simulink models and for the manipulator Testbench.

Let us demonstrate the operation of ICS based on KBO on soft computing with one FC in the conditions of an internal unexpected control situation (*Fig. 37*) in comparison with ICS based on KBO on quantum computing.

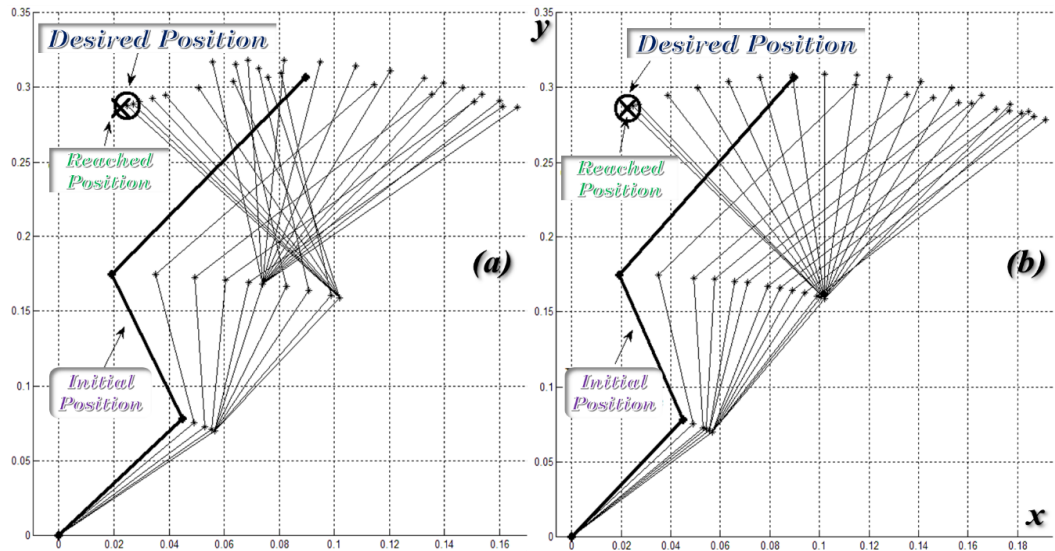

*Figure 37: ICS based on KBO on soft computing with one FC in the conditions of an internal unexpected control situation (a); ICS based on quantum computing (b).*

From *Fig. 37* it can see that ICS based on KBO on quantum computing provides a better solution quality than ICS based on soft computing with one FC.

Consider the control signals generated by the considered types of ICS based on soft and quantum computing.

*4.1.3. Analysis of the control signals in ICS.* In *Fig. 38* *QFC* are the control signals of ICS based on KBO on quantum computing, *FC* are the control signals generated by ICS on soft computing with one FC, *FC Decomposition* are the control signals generated by ICS on soft computing with separated control.

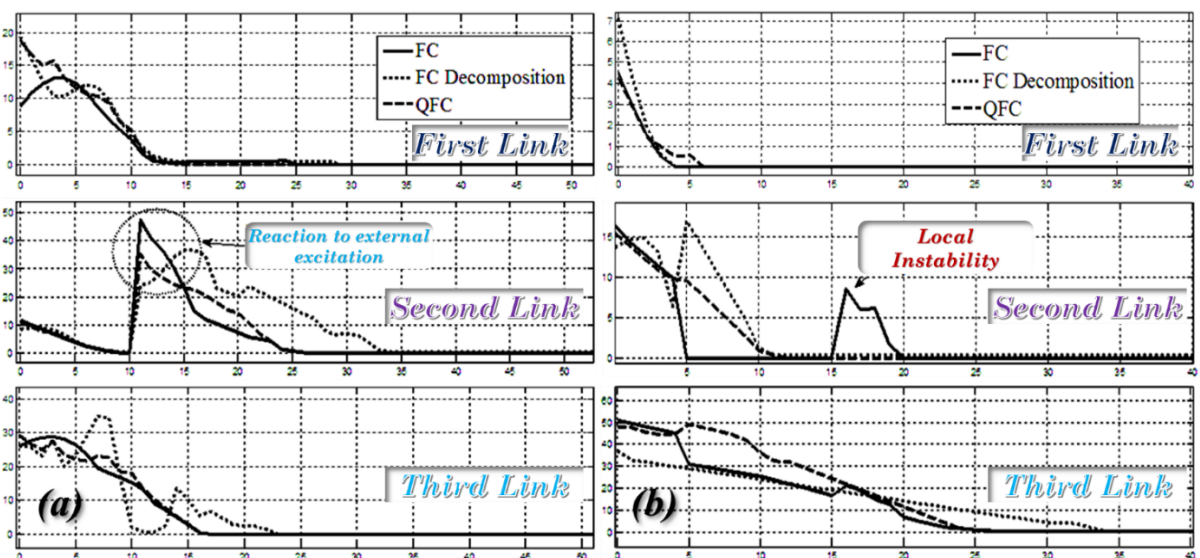

*Figure 38: Control signals generated by different types of ICS in unexpected control situations.*

From *Fig. 38(a)* it can see that the control signals generated by ICS based on KBO on quantum computing under external disturbing influences are similar to those generated by ICS on soft computing with one FC. However, the amplitude of the control signal at the time of external influence is significantly lower for ICS based on quantum computing. The control signal formed by ICS based on soft computing with separated control have become overshoot.

From *Fig. 38(b)* it can see that the control signal generated by ICS based on KBO on soft computing with one FC for the second link has a local instability section, while ICS based on quantum computing generates robust control signals under conditions of external disturbing influences.

*Fig. 39* shows the dynamics of PID controller coefficients at the input and output of the QFI (left and right columns, respectively) for the considered abnormal situation on the example $k_{p1}$.

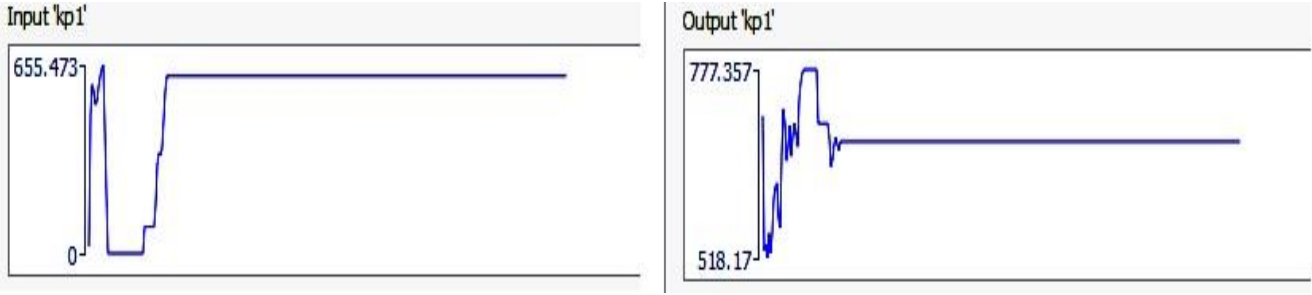

*Figure 39: Dynamics of the PID controller coefficients changes at QFI input and output.*

*Fig. 40* shows a comparison of the control laws for three manipulator links: those generated by soft-computing KBO with split control (thick lines) and quantum-computing KBO (thin lines).

From *Fig. 40* follows that the QFI block inclusion gives the opportunity to obtain control signals adequate to perturbations with less smooth fluctuations, than at the FC output.

The previously considered cases of control contingencies ([1, 8]) - change of initial conditions, forced displacement of links - refer to external control contingencies.

However, in addition to external disturbances, changes in the internal configuration of the object and the control system related to incomplete initial description, disturbances in the control channels, inaccuracy and inertia of the measuring system, etc., are also possible.

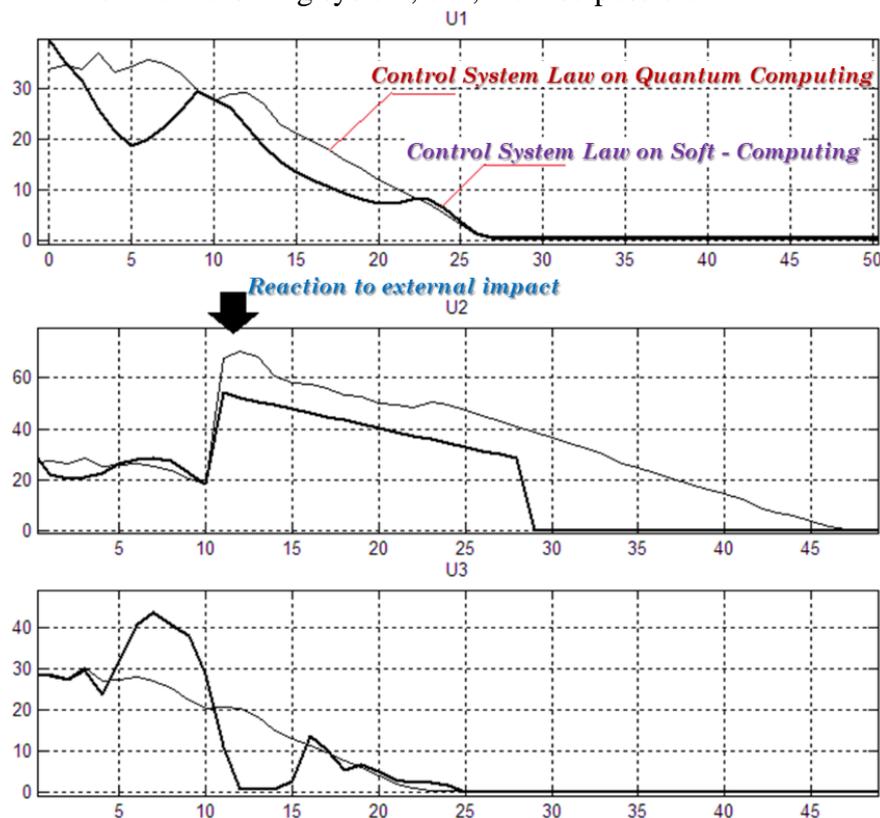

*Figure 40: Control Systems Laws Comparison for Split Control Soft-Computing and Quantum-Computing KBO's.*

Inserting into the control systems models additional cases of control contingencies associated with changes in the parameters of the CO during the time.

## 4.2 Simulation of unexpecting control situations under conditions of CO parameters changes.

Let's define 2 cases of contingency control situations (the initial conditions correspond to normal control situations):

1) at the eleventh iteration the limitation of the output action in terms of changes of links positions changes: it increases from 3 to 5 deg.
2) at the eleventh iteration the output limitation in terms of changes in sections' positions is changed: it decreases from 3 to 1 deg.

Let's compare control laws, generated by ICS on KBO with quantum computing using spatial, spatial-time and time correlations (*Fig. 41*) for contingency control situations in conditions of CO parameters changing.

From *Fig. 41* shows that at control law forming the minimum consumption of useful resource is achieved by using spatial correlation.

For the considered experiments (unexpecting control situations in conditions of CO parameters changing) let's compare control laws, formed by ICS on KBO with quantum computing using spatial correlation, as well as ICS on KBO on soft computing with one FC and split control (*Fig. 42*).

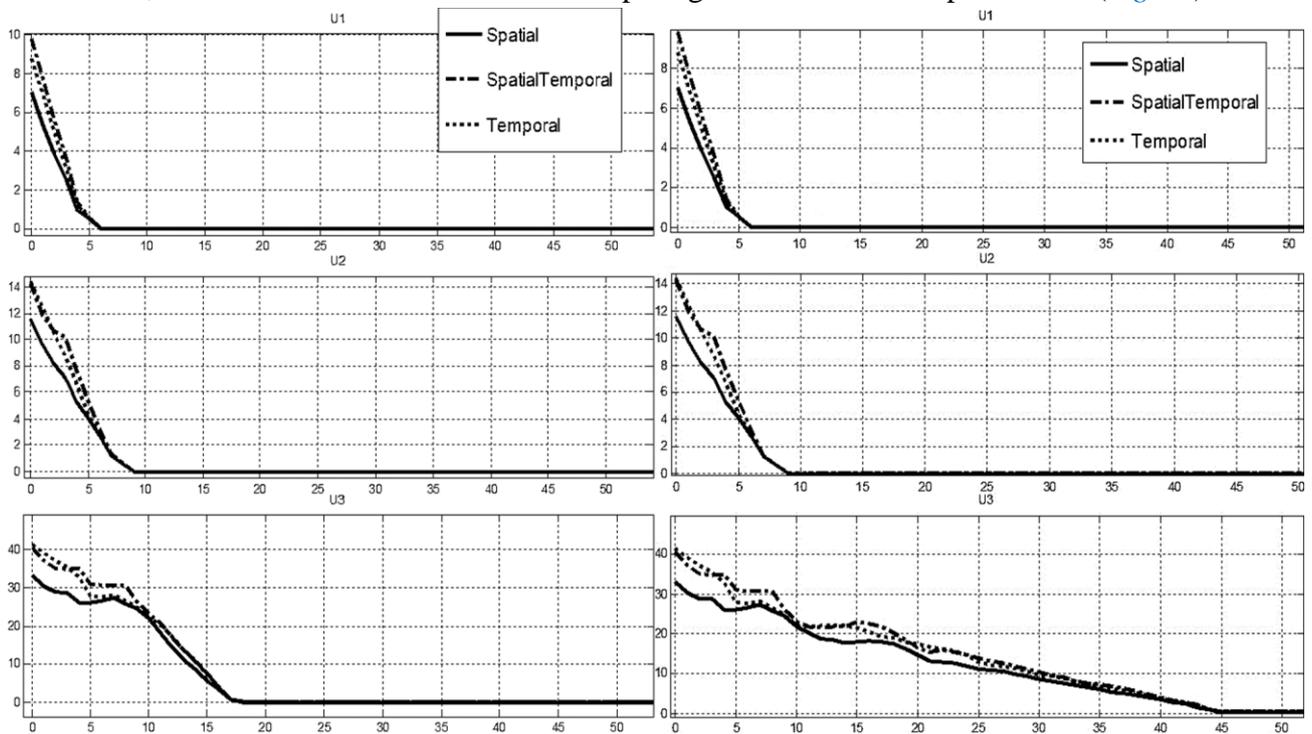

*Figure 41: Comparison of the ICS's formed control laws on KBO with quantum computing using spatio-temporal and temporal correlations in contingent control situations under conditions of the CO parameters changing: first case (left); second case (right).*

For the considered experiments (*Fig. 37*) from the observing point the minimization of the useful resource consumption, the best control laws are formed by ICS with KBO on soft computing with one FC, the insignificant deterioration is observed at formation by ICS with KBO on quantum computing with application of spatial correlation.

The control laws formed by ICS with KBO with split control under the conditions of CO parameters changing has acquired oscillatory character.

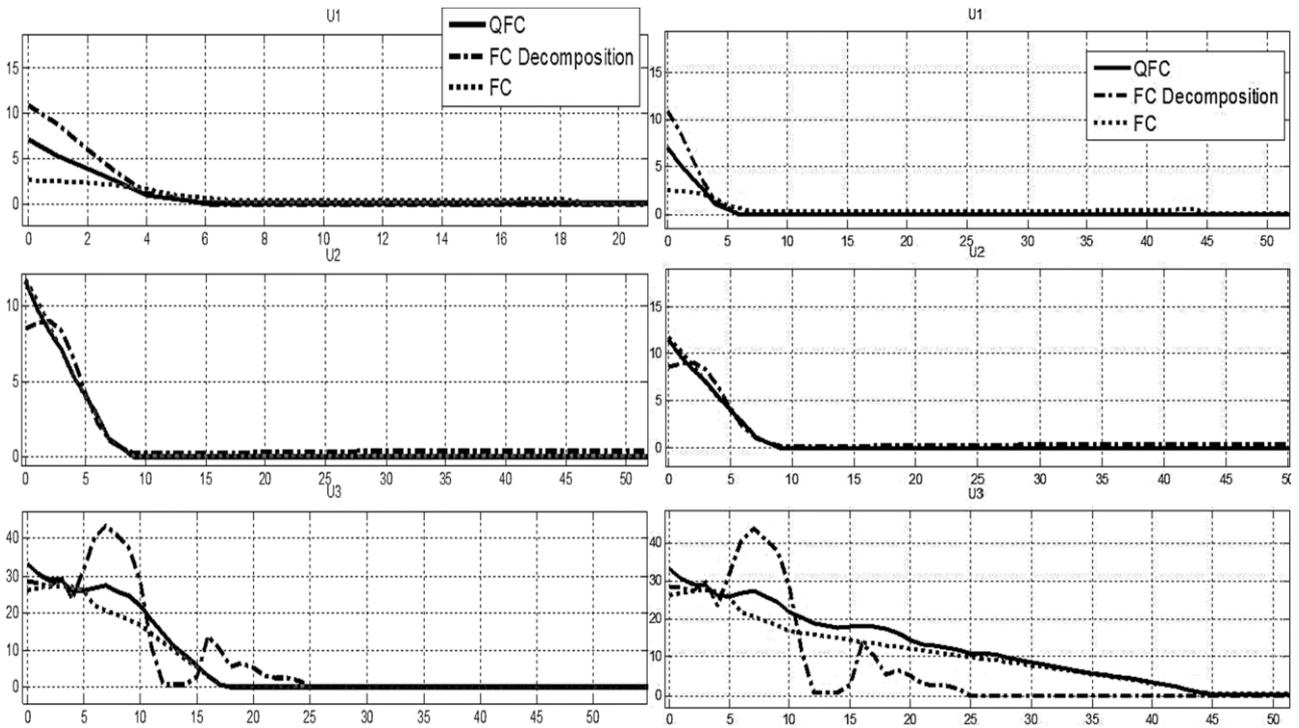

*Figure 42: Comparison of the ICS's formed control laws: ICS on KBO with quantum computing using spatial correlation, ICS on KBO with soft computing using one FC & ICS on KBO with soft computing using split control in contingent control situations under conditions of the CO parameters changing: first case (left); second case (right).*

Let's consider another pair of experiments, which are also corresponded to the both considered cases of unexpected control's situations with the conditions of CO parameters changing of (*Fig. 43*).

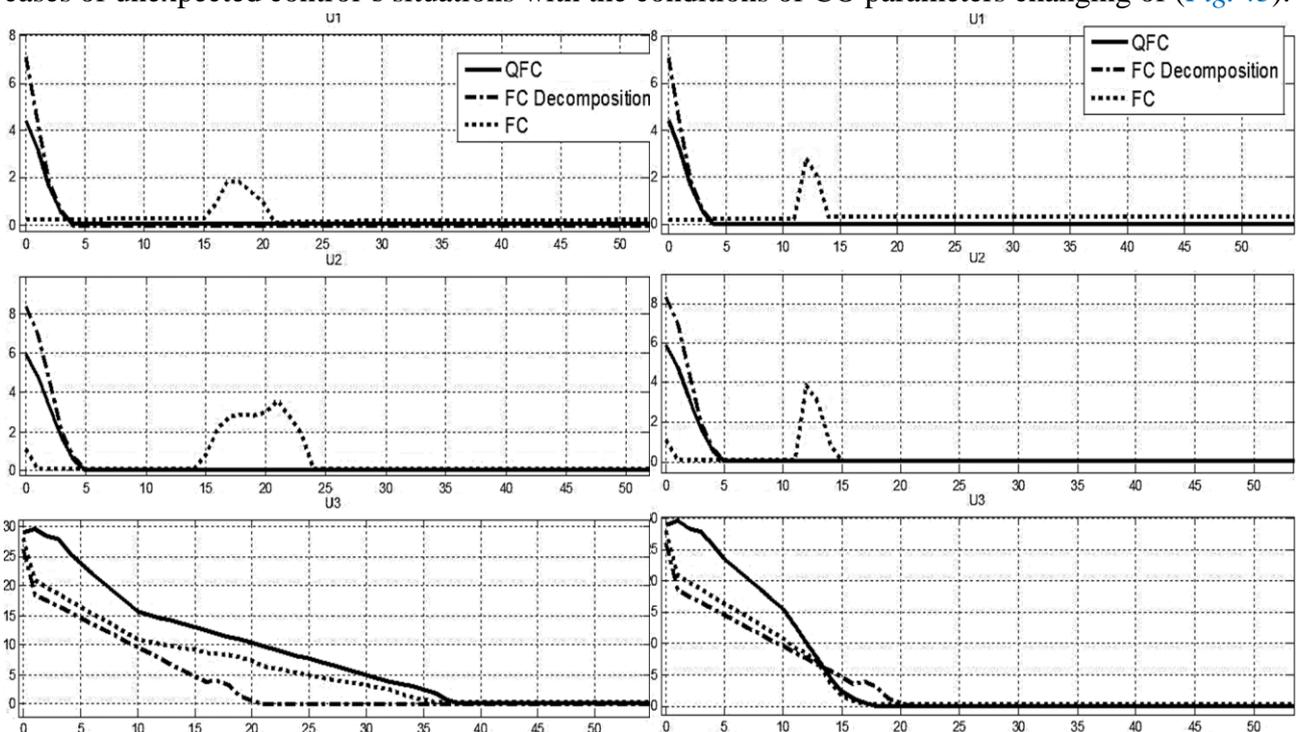

*Figure 43: Comparison of the ICS's formed control laws: ICS on KBO with quantum computing using spatial correlation, ICS on KBO with soft computing using one FC & ICS on KBO with soft computing using split control in contingent control situations under conditions of the CO parameters changing: first case (left); second case (right).*

When considering the experiments (*Fig. 43*), stable control laws were formed by ICS with KBO on quantum computing using spatial correlation and ICS with KBO on soft computing with split control, whereas in the control laws formed by ICS with KBO on soft computing with one FC there are regions of locally unstable states.

Thus, the minimum consumption of useful resource in the formation of control signals ensured when using ICS based on KBO on quantum computing. The fuzzy surface of the parameters of the $K_P$ of the hybrid fuzzy PID controller for the second link, for example, before and after exposure are shown in *Fig. 44*.

This result is especially noteworthy in that when organizing coordination control due to a single knowledge base (correspondingly using one FC in the intelligent control systems for the SCO on soft computing) the number of input variables - i.e. the parameters that determine the functioning of the system are limited by the computing resources of the processor on which the FC is created and the amount of memory in the system in which the FC is located. Moreover, for complex systems, such as a robot manipulator with 7DoF, the organization of a single FC is impossible.

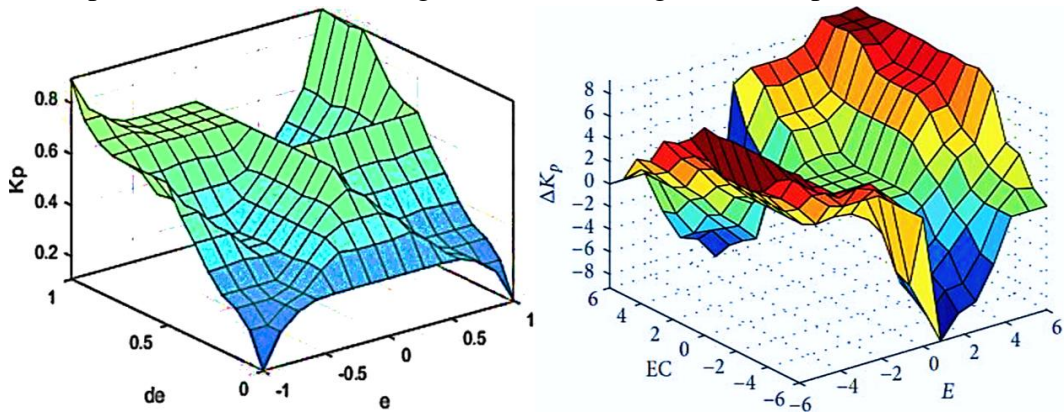

*Figure 44: The fuzzy surface of the parameters of the $K_P$ of the hybrid fuzzy PID controller for the second link before (a) and after (b) exposure.*

In general, the possibility of decomposing the control (dividing one KB into several identical independent KBs) and organizing coordination management by introducing the quantum fuzzy factor link significantly increases the possible number of input variables and thereby expands the possibilities of accounting for the parameters of the system and the control object.

Further research focused on the development and analysis of the physical Testbench of Manipulator with 7 DOF, as well as its integration with the mobile platform. The project Thingiverse was adopted as the basis for creating a prosthesis with a cognitive control system. Servomotors are used as motors. The Arduino controller was used to control the servomotors and control, respectively, the prosthesis itself. To remove the signal of brain activity, the Emotiv EPOC + cognitive helmet was used. The schematic diagram of the device and basic design Steps of hybrid intelligent cognitive control system for prosthetic limb shown of *Fig. 45*.

This example clearly demonstrates the possibilities of sharing electromechanical, anthropomorphic devices, a modern class of computer-based neural interfaces and technologies for intelligent control system design.

Therefore, the extracted information, based on knowledge in the KB of the cognitive controller, makes it possible to obtain an additional information and energy resource for performing useful work, which is equivalent to the appearance of a purposeful action on the control object for guaranteed achievement of the control goal. This fact confirms the validity of the assertion about the physical (thermodynamic) nature of information [8, 10], and the control force using this information makes it possible to perform additional useful work [9, 10].

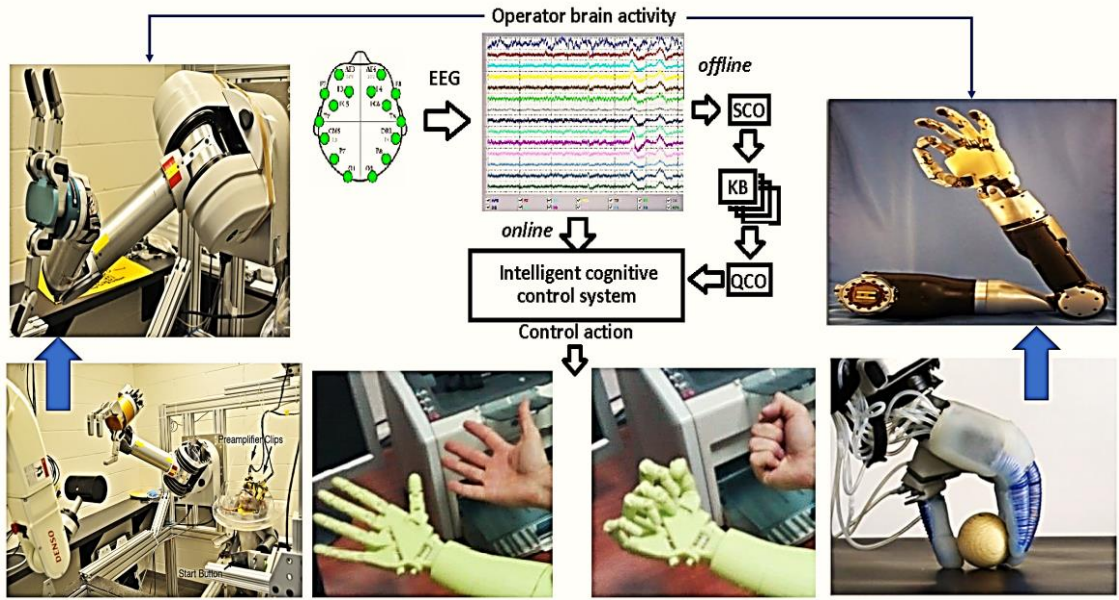

*Figure 45: Design mode (offline) and use mode (online) of intelligent cognitive prosthesis.*

Thus, the thesis "Knowledge is Power" is reformulated in a new design principle: "Cognitive controller is a quantum thermodynamic additional control force" and demonstrate the power of a cognitive intelligent control.

*Figure 46* shows a generalized comparison of ICS based on KBO on soft and quantum computing for 3DOF and 7DOF manipulators for standard and unexpected control situations of the examples considered in this Item).

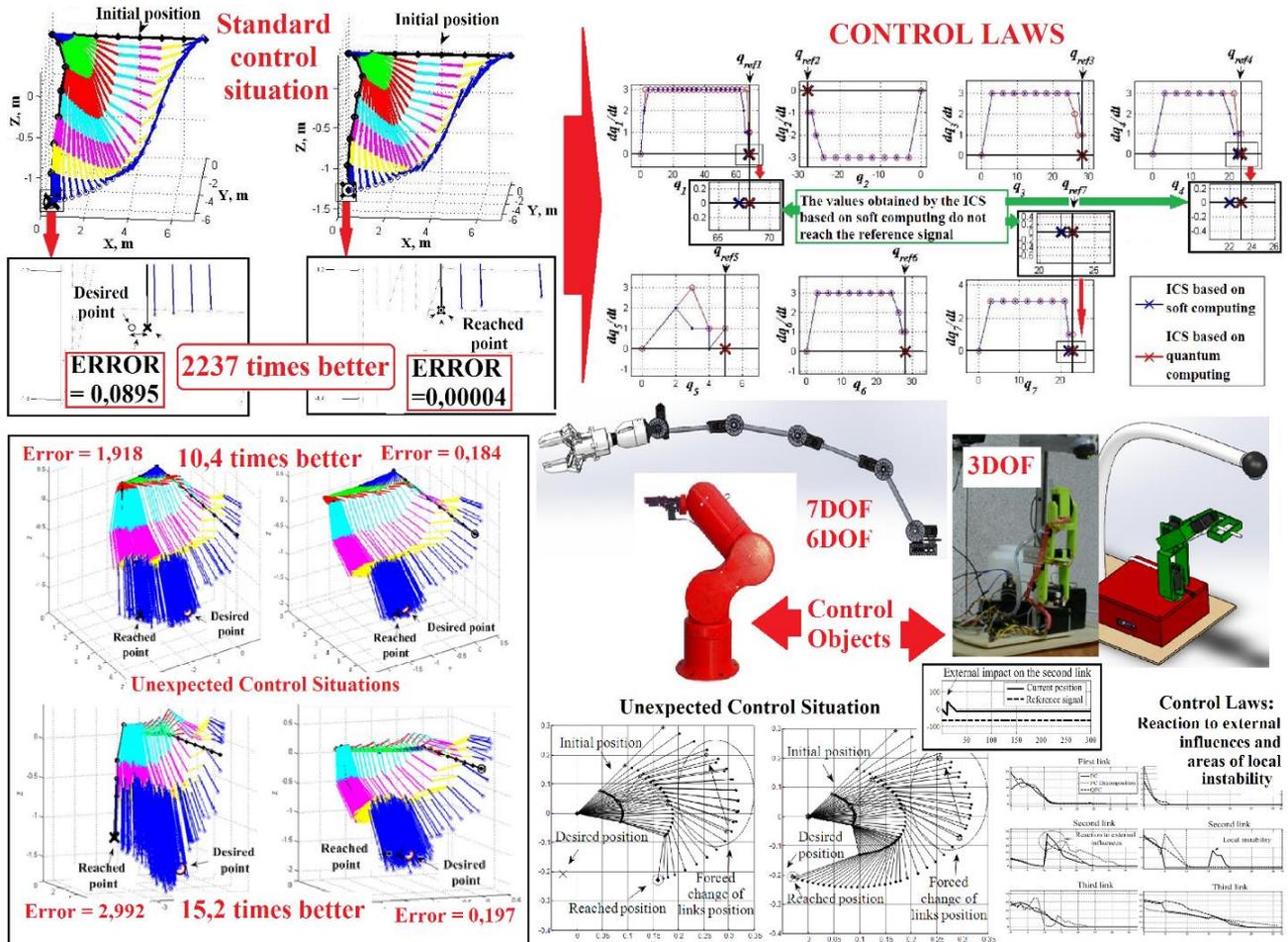

*Figure 46: Comparison of ICS based on soft and quantum computing for 3DOF and 7DOF new unconventional intelligent control type based on quantum fuzzy inference algorithm.*

Using two COs of varying complexity (3DoF and 7DoF manipulators) as an example, the advantages and limitations of using soft and quantum computing technologies in ICS were demonstrated. On the example of 3DOF manipulator, the minimal difference between the results of the physical manipulator Testbench and the MatLab/Simulink model demonstrated.

# Conclusions

- This work presents the development of several high-tech areas of robotics, which have practical scientific and technical interest, both in separate and in joint developments.
- It has been shown that the prospect of developing cognitive intelligent control using soft and quantum computing technologies is one of the important tasks in creating a robotic prosthetic arm, such as a simple case of a robot avatar, and is integral to the development of information technology in the framework of an intelligent simulator concept [8]
- The use of expert recommendation systems with a deep representation of knowledge and quantum end-to-end technologies of deep machine learning with quantum EEG processing allows the appointment, selection of control of robotic prostheses of the hand, taking into account the individual psychophysiological characteristics of the patient and the operating environment.
- On the one hand, these are end products that, if properly developed, can be presented on the market of commercially attractive products; on the other hand, technologies for using new types of intelligent information technologies and human-machine cognitive interfaces.
- The next stage of development is the creation of a cognitive intellectual control system for a robotic arm-prosthesis for maintenance based on IT quantum soft computing, quantum EEG processing filters and Kansei / Affective Engineering intelligent computing technologies with an assessment of the user's emotional state.
- The work, in its essence, reflects the completeness of the formation of a new educational approach in intelligent robotics - a hybrid cognitive intelligent robotics based on neural interfaces with new types of IT data processing.